\documentclass[twocolumns]{aa}  

\usepackage{graphicx}
\usepackage{txfonts}
\usepackage{hyperref}

\usepackage{listings} 
\usepackage{xcolor} 

\hypersetup{
    colorlinks=true,
    citecolor=blue,
    linkcolor=blue,
    urlcolor=blue,
}

\expandafter\def\expandafter\UrlBreaks\expandafter{\UrlBreaks
  \do\a\do\b\do\c\do\d\do\e\do\f\do\g\do\h\do\i\do\j%
  \do\k\do\l\do\m\do\n\do\o\do\p\do\q\do\r\do\s\do\t%
  \do\u\do\v\do\w\do\x\do\y\do\z\do\A\do\B\do\C\do\D%
  \do\E\do\F\do\G\do\H\do\I\do\J\do\K\do\L\do\M\do\N%
  \do\O\do\P\do\Q\do\R\do\S\do\T\do\U\do\V\do\W\do\X%
  \do\Y\do\Z}

\makeatletter
\renewcommand*\aa@pageof{, page \thepage{} of \pageref*{LastPage}}
\makeatother

\addto\extrasenglish{%
}

\definecolor{WHITE}{RGB}{255, 255, 255} 
\definecolor{codeblackkey}{RGB}{46, 46, 46}
\definecolor{codeblackid}{RGB}{51, 51, 51}
\definecolor{codegreycom}{RGB}{153, 153, 136}
\definecolor{codegreyfrm}{RGB}{153, 153, 153}
\definecolor{codegreennum}{RGB}{0, 153, 153}
\definecolor{coderedstr}{RGB}{221, 17, 68}
\definecolor{codered}{RGB}{192, 52, 29}
\definecolor{coderedhl}{RGB}{251, 229, 255}

\newcommand*{\FormatDigit}[1]{\textcolor{codegreennum}{#1}} 
\newcommand*{\FormatBool}[1]{\textcolor{codegreycom}{#1}} 
\newcommand*{\FormatBase}[1]{\textcolor{black}{#1}} 

\begin{document}

   \title{Four HD~209458~b transits through CRIRES+: Detection of H$_2$O and non-detections of C$_2$H$_2$, CH$_4$, and HCN}

   \author{
        D. Blain\inst{1}
        \and
        R. Landman\inst{2}
        \and
        P. Mollière\inst{1}
        \and
        J. Dittmann\inst{3}
    }

   \institute{
        Max-Planck-Institut für Astronomie, Heidelberg, Germany\\
        \email{blain@mpia-hd.mpg.de}
        \and
        Leiden Observatory, Leiden University, Leiden, The Netherlands\\
        \and
        University of Florida, Gainesville, United States of America
    }

   \date{Received 17 May 2024 / Accepted DD MM 202x}

 
  \abstract
   {HD~209458~b is one of the most studied exoplanets to date. Despite this, atmospheric characterisation studies yielded inconsistent species detections and abundances. Values reported for the C/O ratio range from $\approx$ 0.1 to 1.0. Of particular interest is the simultaneous detection of H$_2$O and HCN reported by some studies using high-resolution ground-based observations, which would require the atmospheric C/O ratio to be fine-tuned to a narrow interval around 1. HCN has however not been detected from recent space-based observations.}
   {We aim to provide an independent study of HD~209458~b's atmosphere with high-resolution observations, in order to infer the presence of several species, including H$_2$O and HCN.}
   {We observed four primary transits of HD~209458~b at a high resolution ($\mathcal{R} \approx 92\,000$) with CRIRES+ in the near infrared (band H, 1.431243--1.837253 $\mu$m). After reducing the data with \texttt{pycrires}, we prepared the data using the SysRem algorithm and performed a cross-correlation (CCF) analysis of the transmission spectra. We also compared the results with those obtained from simulated datasets constructed by combining the Exo-REM self-consistent model with the petitRADTRANS package.}
   {Combining the four transits, we detect H$_2$O with a signal-to-noise CCF metric of $8.7 \sigma$. This corresponds to a signal emitted at $K_p = 151.3^{+31.1}_{-23.4}$ km$\cdot$s$^{-1}$ and blueshifted by $-6^{+1}_{-2}$ km$\cdot$s$^{-1}$, consistent with what is expected for HD~209458~b. We do not detect any other species among C$_2$H$_2$, CH$_4$, CO, CO$_2$, H$_2$S, HCN, and NH$_3$. Comparing this with our simulated datasets, this result is consistent with a C/O ratio of 0.1 and an opaque cloud top pressure of 50 Pa, at a 3 times solar metallicity. This would also be consistent with recent \textit{JWST} observations. However, none of the simulated results obtained with a bulk C/O ratio of 0.8, a value suggested by previous studies using GIANO-B and CRIRES, are consistent with our observations.}
   {}


   \maketitle

\section{Introduction}
    \label{sec:introduction}

    \begin{figure}
       \centering
       \includegraphics[width=\hsize]{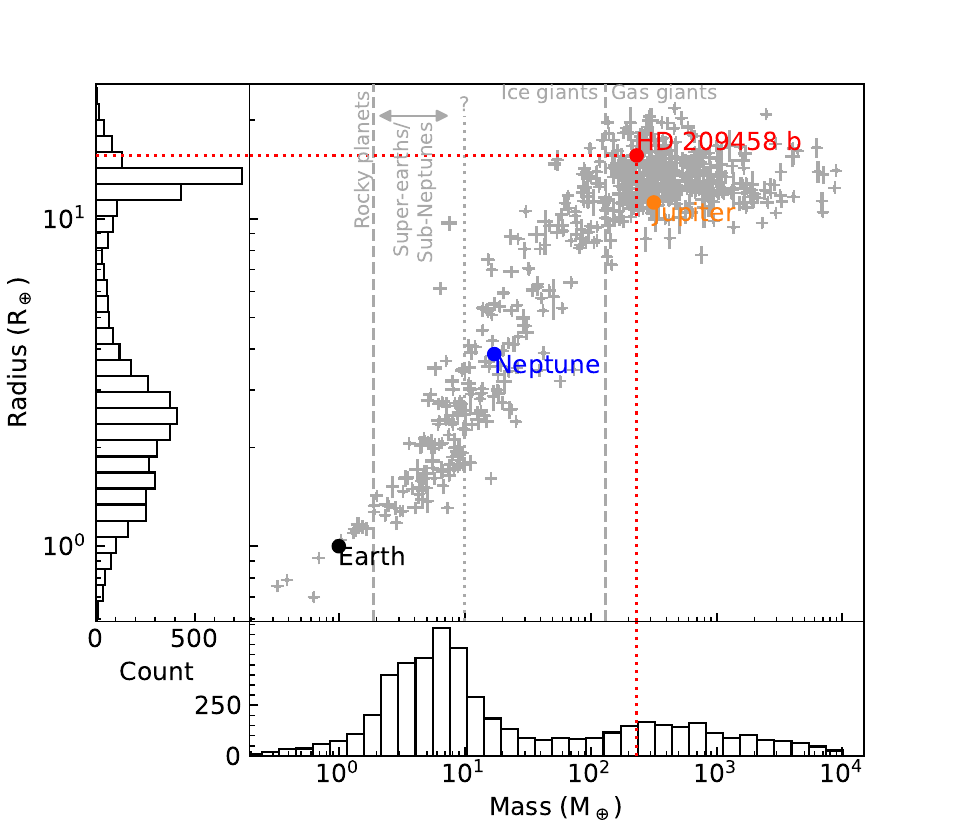}
          \caption{
            Mass-radius distribution of the 5630 confirmed exoplanets to date \citep[17 May 2024,][]{NasaExoplanetArchive}. Not all the confirmed planets have a measured radius and/or mass, and thus the total count in the histograms is less than 5630. The scatter plot shows only the 596 planets for which the mass and radius are known within 15$\%$. 
            }
        \label{fig:exoplanet_distribution}
    \end{figure}

    Valuable information can be extracted from the characterisation of exoplanet atmospheres, such as dynamics, aerosol presence, and chemical composition. The latter can be a key element in the understanding of a planet's history and formation process \citep[e.g.][]{Oberg2011, Madhusudhan2014b, Mordasini2016, Molliere2022}.
    
    One of the currently most studied parameters in planet's atmospheres is the carbon-to-oxygen (C/O) ratio, which is of particular interest to understand planetary formation histories. For gas giants, a wide variety of formation scenarios are proposed to explain both potentially O-rich or C-rich bulk compositions compared to the solar C/O ratio \citep[$\approx 0.55$, see][]{Lodders2019}. For O-rich compositions (i.e. $\rm C/O \lesssim$~stellar), scenarios include the accretion of silicates and water ice or clathrates through planetesimals or pebbbles, or the inward drift and evaporation of pebbles rich in O-dominated volatile ices (mostly water) \citep[e.g.][]{Gautier2005, Monga2015, Mordasini2016, Booth2017}. C-rich composition scenarios (i.e. $\rm C/O \gtrsim$~stellar) favour the involvement of tar, enrichment dominated by gas accretion, and evaporating pebbles that transport C-rich ices from the outer disk, or they could be explained by a high C/O ratio in the protostellar nebula \citep[e.g.][]{Lodders2004, Oberg2011, Booth2017, Pekmezci2019}. This variety of scenarios arises from the fact that the C/O ratios of the most studied gas giants, those in the Solar System, are difficult to estimate due to several factors, including the condensation of H$_2$O clouds deep in the atmosphere of these objects \citep[see e.g.][]{Cavalie2024}. Most of the currently detected exo-gas-giants have an equilibrium temperature too high for H$_2$O condensation. Hence, in principle, the C/O ratio of these planet's atmosphere should be easier to determine, making for an interesting comparative study with the Solar System's gas giants.

    With a radius $\approx 40\%$ larger than that of Jupiter while having $\approx 30\%$ less mass (see \autoref{fig:exoplanet_distribution}), its equilibrium temperature of $1450$ K and its bright host star, HD~209458~b, is one of the best targets for atmospheric characterisation. These properties give this planet the highest transmission spectroscopy metric \citep[TSM,][]{Kempton2018} of all known planets, $\approx 900$. Consequently, this exoplanet is one of the most studied to date.

    Multiple species have been detected in HD~209458~b's atmosphere. At wavelengths shorter than the infrared, several atomic and ionic species detections (e.g. Lyman-$\alpha$ and He I) or tentative detections (e.g. Ca I, Mn I, and Sc II) have been reported \citep[e.g. recently, ][]{Alonso2019b, Cubillos2020, Lira2022}. In the infrared, in addition to CO \citep{Brogi2017, Brogi2019, Gandhi2019} and H$_2$O \citep{Beaulieu2010, Deming2013, Madhusudhan2014, Brogi2017, Gandhi2019, Sanchez2019}, with the detection of HCN was reported by \citet{Hawker2018, Giacobbe2021}, the latter study also reporting the detection of C$_2$H$_2$, CH$_4$, HCN, and NH$_3$. In contrast, \citet{Xue2024} detected only H$_2$O and CO$_2$\footnote{The data taken by \citet{Xue2024} contain the strong CO$_2$  $v_3$ band at $\approx$ 4 $\mu$m \citep[see e.g.][]{Yurchenko2020}. However, CO$_2$ does not have significant spectral features in band H, hence it cannot be detected from our dataset (see \autoref{subsec:self_consistent_model} and \autoref{subsubsec:simulated_data_results}).}, and provided upper limits for C$_2$H$_2$, CH$_4$, HCN, and NH$_3$. Other studies, such as \citet{Schwarz2015}, reported a non-detection of H$_2$O and CO.

    Some of these detections are, however, conflicting. \citet{Xue2024} inferred a sub-solar C/O ratio ($\approx 0.1$), while the detection of HCN in HD~209458~b's atmosphere would suggest instead a super-solar C/O ratio ($\approx 1$). Moreover, the simultaneous detection of HCN and H$_2$O would only be allowed by a narrow range of C/O ratios, according to equilibrium chemistry. It could also be the sign of silicate clouds condensing on the nightside, and evaporating on the dayside, creating a chemical asymmetry between the morning and evening terminator, for which a wider range of bulk C/O ratios ($\sim$ 0.7-0.9) may be allowed \citep[][]{Sanchez2022}. In this scenario the morning terminator could have a gas-phase C/O~$\gtrsim 1$, due to the sequestration of oxygen into silicates, while the evening terminator would have a gas-phase C/O~$\lesssim 1$. Interestingly, both HCN detections for HD~209458b were claimed from high-resolution ($\mathcal{R} \approx 100\,000$) ground-based observations, while the result from \citet{Xue2024} comes from low-resolution ($\mathcal{R} \approx 3\,000$) space-based -- James Webb Space Telescope, \textit{JWST} -- data. 
    
    This inconsistency is one of several currently involving exoplanet characterisation. For HD~209458~b, concurrent studies also claim incompatible H$_2$O volume mixing ratios, from $\approx 10^{-6}$ to $10^{-2}$ \citep[e.g.][]{Madhusudhan2014, Brogi2017, MacDonald2017, Tsiaras2018, Gandhi2019, Xue2024}, while studies on high-resolution datasets report inconsistent spectral feature Doppler-shifting, from $\approx -6$ km$\cdot$s$^{-1}$ to 0 km$\cdot$s$^{-1}$ \citep[e.g.][]{Snellen2010, Gandhi2019, Sanchez2019, Giacobbe2021}.

    Given this context, the photometric band H ($\approx 1.3$--$1.9$ $\mu$m) is of high interest, as it contains spectral bands of H$_2$O and HCN, but also C$_2$H$_2$, CH$_4$, CO, H$_2$S, and NH$_3$. This makes this band an excellent candidate to infer the C/O ratio, but also the N/H and S/H ratios.

    With this work, we propose an independent analysis of the atmosphere of HD~209458b, making use of four primary transits at a high-resolution in band H, supported by 1D self-consistent models of the atmosphere. We performed a cross-correlation function (CCF) analysis to provide detection significance of multiple species. We compared these results with those obtained from simulated data with different chemical compositions and cloud altitudes.
    
    \begin{table*}[ht]
    \centering
    \caption[]{\label{tab:general_parameters} Parameters of HD~209458~b and its star.}
        \begin{tabular}{@{}lccc@{}} 
        \hline \hline
        Parameter                                           & Value                                                     &                               & References    \\ \hline
        \textbf{Host star:} \\
        Spectral type                                       & G0 V                                                      &                               & 1             \\
        $M_\ast$ (kg)                                       & 2.45 $\pm$ 0.18 $\times 10^{30}$                          & (1.23 M$_\odot$)              & 2             \\
        $R_\ast$ (Gm)                                       & 0.83 $\pm$ 0.01                                           & (1.19 R$_\odot$)              & 2             \\
        $T_{\ast,\,\text{eff}}$ (K)                         & 6091 $\pm$ 10                                             &                               & 2             \\
        $g_\ast$ (m$\cdot$s$^{-2}$)                         & 282 $\pm$ 13                                              & (4.45 [cm$\cdot$s$^{-2}$])    & 2             \\
        $[$Fe/H$]$                                          & 0.01                                                      &                               & 2             \\
        $t_{\ast}$ (Gyr)                                    & 3.10$^{+0.80}_{-0.70}$                                    &                               & 3             \\
        $V_{\mathrm{sys}}$ (km$\cdot$s$^{-2}$)              & -14.741 $\pm$ 0.002                                       &                               & 4             \\
        RA/Dec (J2000, epoch 2015.5)                        & $330.795022^\circ\,18.884242^\circ$                       & (22:03:10.81 +18:53:03.27)    & 5             \\
        \textbf{Planet:} \\
        $a_p$ (Gm)                                          & 7.29 $\pm$ 0.15                                           & (0.0488 au)                   & 2             \\
        $e$                                                 & $0.0$                                                     &                               & 2             \\
        $i_p$ (degree)                                      & 86.71$\pm$ 0.05                                           &                               & 2             \\
        $M_p$ (kg)                                          & 1.39 $\pm$ 0.02 $\times 10^{27}$                          & (0.73 M$_J$)                  & 2             \\
        $R_p$ (km)                                          & 99$\,$500 $\pm$ 1300                                      & (1.39 R$_J$)                  & 2             \\
        $g_p$ (m$\cdot$s$^{-2}$)                            & 9.36 $\pm$ 0.58                                           & (2.97 [cm$\cdot$s$^{-2}$])    & Derived       \\
        $T_{\text{eq}}$ (K)                                 & 1451 $\pm$ 20                                             &                               & Derived       \\
        $T_{p,\,\text{int}}$ (K)                            & $\approx$ 550                                             &                               & 6 (model)     \\
        $P$ (day)                                           & 3.5247404585539 $\pm$ 0.000015326508                      &                               & 5             \\
        $T_{14}$ (s)                                        & 11002.198566627 $\pm$ 1.324210291                         &                               & 5             \\
        \textbf{Mid-transit times:} \\
        $T_{0,1}$ (BJD$_{\text{TDB}}$, day)                 & 2459759.810949 $\pm$ 0.000298                             &                               & Derived       \\
        $T_{0,2}$ (BJD$_{\text{TDB}}$, day)                 & 2459766.860430 $\pm$ 0.000268                             &                               & Derived       \\
        $T_{0,3}$ (BJD$_{\text{TDB}}$, day)                 & 2459865.553163 $\pm$ 0.000181                             &                               & Derived       \\
        $T_{0,4}$ (BJD$_{\text{TDB}}$, day)                 & 2460122.859217 $\pm$ 0.001289                             &                               & Derived       \\ \hline
        \end{tabular}
    \tablebib{
    (1)~\citet{delBrugo2016}; (2) \citet{Stassun2017}; (3) \citet{Bonomo2017}; (4) \citet{Naef2004}; (5) \citet{exofopweb}; (6) \citet{Thorngren2020}. Some parameters were acquired from the \citet{NasaExoplanetArchive}.
    }
    \end{table*}

\section{Observations}
    \label{sec:observations}
    
    We observed four transits of HD~209458b in H-band with CRIRES+ \citep{Follert2014} at the Very Large Telescope (VLT) as part of the ESO proposal 109.2376 (PI: Molli\`ere). Obervations were taken in the nights of 29 June 2022, 6 July 2022, 12 October 2022, and 27 June 2023, called Night 1--4 in the following. We used a reference wavelength of 1567.099~nm and a slit width of 0.2'' for all nights, corresponding to spectral resolutions $\gtrsim 100,000$. The observations were performed with the adaptive optics (AO) system MACAO \citep{Arsenault2003}, using the science target, HD 209458, as the AO star, allowing us to use the 0.2'' slit efficiently despite the larger seeing conditions. During nights 1--3, observations were taken in nodding mode with a 6'' nod throw. For Night 1 and 2 a detector integration time of ${\rm DIT}=30$~s was used, with six exposures per nodding position. For Night 3 this was changed to 3~exposures per nodding position. For Night 4 nodding was turned off, with a DIT of 20~s. Our observations were set up to avoid airmasses larger than 2, with total exposure times to cover the full transit, and to establish significant out-of-transit baseline observations. Typical execution times were of the order of 6~h. This results in 468, 600, 474, and 838 exposures taken for nights 1 to 4, respectively. The seeing was below 0.5'' during Night~1, between 0.5 and 1'' during Night~2, between 1 and 1.5'' during Night~3 and between 0.5 and 1.5'' during Night~4. The precipitable water vapors (PWV) were 1.5--1.6 mm, 1.5--2.5 mm, 1.5--1.7 mm, and 2.4--4.0 mm during nights 1 to 4, respectively.

    
    The data were reduced using \texttt{pycrires} \citep{Stolker2023}, which is a python wrapper for the official ESO CR2RES pipeline (Version 1.4.0)\footnote{\url{https://ftp.eso.org/pub/dfs/pipelines/instruments/cr2res/cr2re-pipeline-manual-1.4.0.pdf}}. This pipeline includes the basic calibration steps, including dark and flat field correction, and optimally extracts the stellar spectrum. For the observations that were taken in nodding mode, we subtracted the background using an observation in the other nodding position. The Esorex pipeline by default combines all the exposures in the same nodding position, decreasing the temporal resolution. To avoid this, we run the pipeline on each nodding pair separately. We used each exposure only once, meaning that the first A exposure in a sequence was combined with the first B exposure, the second A exposure with the second B exposure, and so forth. In the spectral extraction we used an extraction height of 20 pixels, a swath width of 400 pixels, and an extraction oversampling of 10. Additionally, we turned off the \texttt{subtract\_no\_light\_rows}, \texttt{subtract\_interorder\_column} and cosmic ray correction, as we found that these provided spurious values of certain spectral bins in the extracted spectrum. The wavelength solution was obtained using the Esorex pipeline and the calibration data from the Uranium-Neon lamp and the Fabry-Pérot etalon. However, we found slight inaccuracies in this wavelength solution by visually comparing it with a telluric model generated using SkyCalc. With CRIRES+, excellent seeing conditions ($\lessapprox 1$'') can lead the 0.2'' slit to receive a significantly non-uniform flux along its width, in an effect called 'super-resolution'. In the case the position of the target on the slit changes with exposures, this can have a significant effect on the lines position ($\approx 0.01$ nm) and on the line spread function, while increasing the effective resolving power of the instrument. We therefore performed an additional correction on the wavelength solution using the \texttt{correct\_wavelengths} method in \texttt{pycrires}, which is explained in \citep{Landman2024}. An offset and linear correction to the wavelength solution from the Esorex pipeline were obtained by maximising the cross-correlation between the observed spectrum and a telluric model. We found that only small corrections to the original wavelength solution of $<0.01$ nm were required for this wavelength setting. The wavelength solution was subsequently verified through visual comparison with a telluric model.

\section{Models}
    \label{sec:models}

    \subsection{Self-consistent model}
        \label{subsec:self_consistent_model}

        \begin{figure}
           \centering
           \includegraphics[width=\hsize]{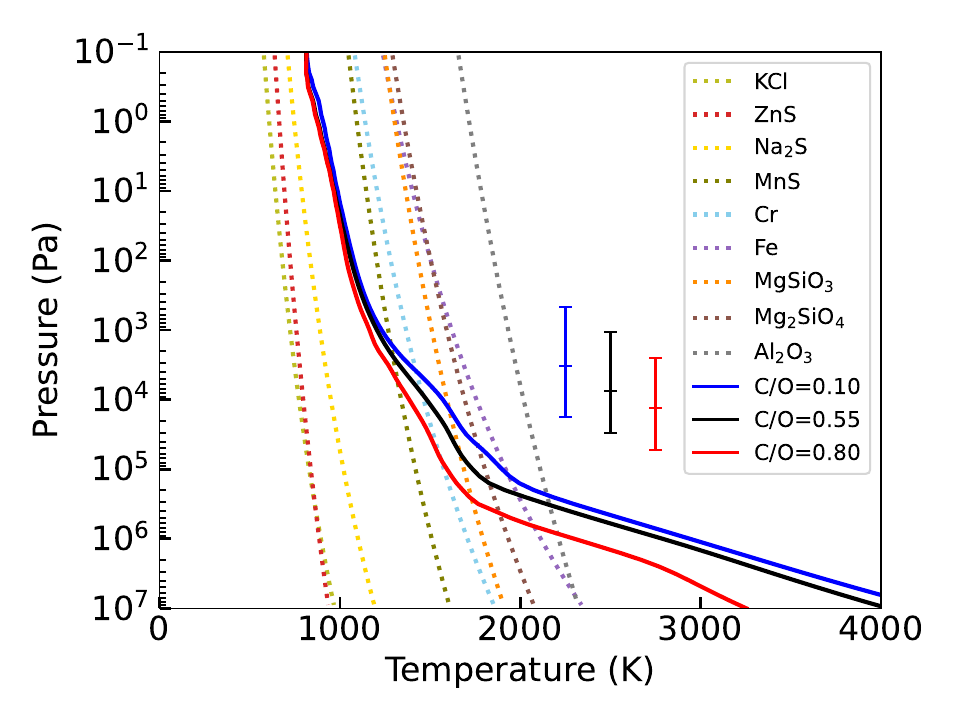}
              \caption{
                Temperature profile and condensation curves obtained from our self-consistent model, for 3 times the solar metallicity. 
                The blue, black, and red errorbars represent the pressure range concentrating $68\%$ of the transmission contribution of our petitRADTRANS simulated data (see \autoref{subsubsec:simulated_data}) over the data wavelength range, for a C/O ratio of 0.1, 0.55, and 0.8, respectively.
                }
            \label{fig:exorem_temperature_profiles}
        \end{figure}

        \begin{figure}
           \centering
           \includegraphics[width=\hsize]{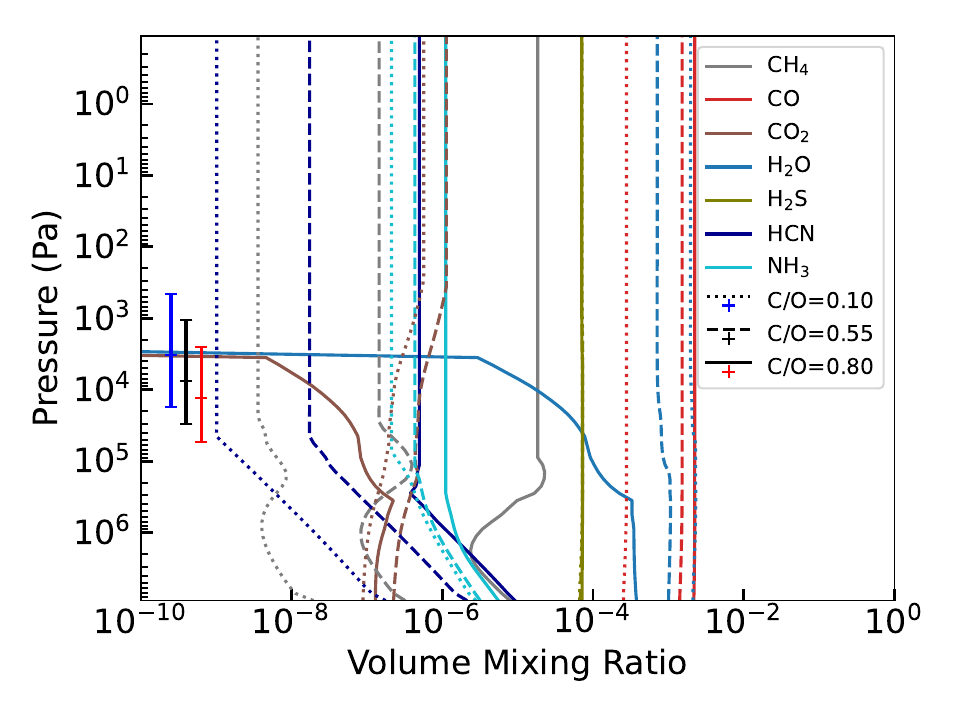}
              \caption{
                Volume mixing ratios obtained from our self-consistent model, for 3 times the solar metallicity. Dotted: $\mathrm{C/O} = 0.1$. Dashed: $\mathrm{C/O} = 0.55$. Solid: $\mathrm{C/O} = 0.8$.  Only the most relevant absorbing species are represented. Na, K and PH$_3$ are not represented due to their negligible contribution in our data wavelength coverage, while FeH, TiO and VO have VMRs lower than $10^{-10}$ at the pressure of maximum sensitivity. The blue, black, and red errorbars represent the pressure range concentrating $68\%$ of the transmission contribution of our petitRADTRANS simulated data (see \autoref{subsubsec:simulated_data}) over the data wavelength range, for a C/O ratio of 0.1, 0.55, and 0.8, respectively.
                }
            \label{fig:exorem_volume_mixing_ratios}
        \end{figure}

        In order to have a robust first estimation of HD~209458~b's atmospheric properties, we generate several self-consistent models using Exo-REM\footnote{{\sloppy \url{https://gitlab.obspm.fr/Exoplanet-Atmospheres-LESIA/exorem}}}. A summary of Exo-REM is provided in \citet{Blain2024}, which is fully described by \citet{Baudino2015, Baudino2017, Charnay2018, Blain2021}.
        
        Our HD~209458~b self-consistent models were parameterised following the same procedure described in \citet{Blain2021}. The planet and star parameters used are displayed in \autoref{tab:general_parameters}. 
        
        HD~209458~b has a radius larger than expected by 'baseline' evolutionary models \citep[e.g.][]{Bodenheimer2001, Guillot2002}. As such, HD~209458~b falls into the category of 'inflated' planets. These larger radii has been attributed to hotter intrinsic temperatures than otherwise expected \citep{Thorngren2020}. Several mechanisms has been proposed to explain these hot interiors, including Ohmic dissipation \citep{Batygin2010, Laughlin2011}, kinetic dissipation \citep{Guillot2002}, or tidal dissipation \citep{Bodenheimer2001}. The latter however may be ruled out for HD~209458~b's, due to the low eccentricity of the planet and apparent absence of another massive planet orbiting the star \citep{Guillot2002}. In any case, because HD~209458~b is inflated, we use the model from \citet{Thorngren2020} to estimate its intrinsic temperature ($T_{p,\mathrm{int}}$).
        
        We include the radiative coupling of Fe, Mg$_2$SiO$_4$, and SiO$_2$ clouds. We chose these condensates because their respective forming gases are the most abundant at the clouds pressures and temperatures of formation, according to our model. We use a cloud fraction of 0.8, which we estimate is reasonable, on average, in regard to 3-D simulations of the planet \citep[e.g.][]{Lines2018}, and is also consistent with the value inferred by \citet{Xue2024}. We use the wavelength- and particle radius-dependent optical constants (extinction coefficient, single-scattering albedo and asymmetry parameter) of Fe and Mg$_2$SiO$_4$ provided by Exo-REM. For SiO$_2$, we use the refractive indices of \citet{Zeidler2013}\footnote{Acquired from the \citet{DOCCD2024}, SiO$_2$ at 928 K, $\mathrm{E}\parallel\mathrm{c}$.}, and converted them to optical constants using the \texttt{optpropgen} code\footnote{Code developed by F. Montmessin and J.-B. Madeleine at LATMOS/LMD, and available on the Exo-REM reporitory.}. We use Exo-REM with the cloud model of \citet{Charnay2018} to self-consistently calculate the radius of the clouds particles, using a sticking efficiency of 1 and with a supersaturation parameter of $3\times10^{-3}$. 
        
        We chose to use an atmospheric metallicity ($Z$) of 3 times the solar metallicity, which is the median value reported by \citet{Xue2024}, and is consistent with the mass-metallicity trend observed in the Solar System and exoplanets in general \citep[see e.g.][]{Thorngren2016, Welbanks2019, Blain2021}. We then compute three models with different C/O ratios, by varying the elemental abundance of C:
        \begin{enumerate}
            \item C/O $= 0.10$: to match the median value reported by \citet{Xue2024},
            \item C/O $= 0.55$: the solar C/O ratio \citep{Lodders2019},
            \item C/O $= 0.80$: close to the C/O ratio reported by \citet{Giacobbe2021}.
        \end{enumerate}

        These C/O ratios above represent the bulk elemental abundances in the deep atmosphere, but due to cloud condensation, the effective C/O ratio in the upper atmosphere may be different. For example, our third model has an effective C/O ratio above the cloud condensation levels ($\approx 10^{3}$ Pa) of $\approx 1$. 
        
        We were not able to obtain Exo-REM models at the same bulk C/O ratio as the best fit from \citet{Giacobbe2021} ($\approx 0.95$) due to the removal of H$_2$O under these conditions. Since the Exo-REM chemical model base most of its calculations on the H$_2$O abundance, its removal from the atmosphere makes it unstable. This low H$_2$O abundance is due to a combination of factors. First, as the abundance of C increases, the atmospheric conditions of HD~209458~b favours the formation of CO at the detriment of H$_2$O. At the same time, H$_2$O is directly involved in the formation of Si-bearing condensates (Mg$_2$SiO$_4$, MgSiO$_3$, SiO$_2$). With a solar composition, silicon is ten times less abundant than oxygen \citep{Lodders2019}. Hence, the condensations of Si-bearing species can efficiently remove H$_2$O from the atmosphere if its abundance is sufficiently low. With most of the oxygen trapped in CO and Si-bearing condensates, the H$_2$O volume mixing ratio can drop much below $10^{-10}$ above the SiO$_2$ cloud formation level. On a side note, this might also prevent the complete removal of SiO from the upper atmosphere, making this species potentially detectable in that scenario.
        
\begin{figure*}[!ht]
           \centering
           \includegraphics[width=\hsize]{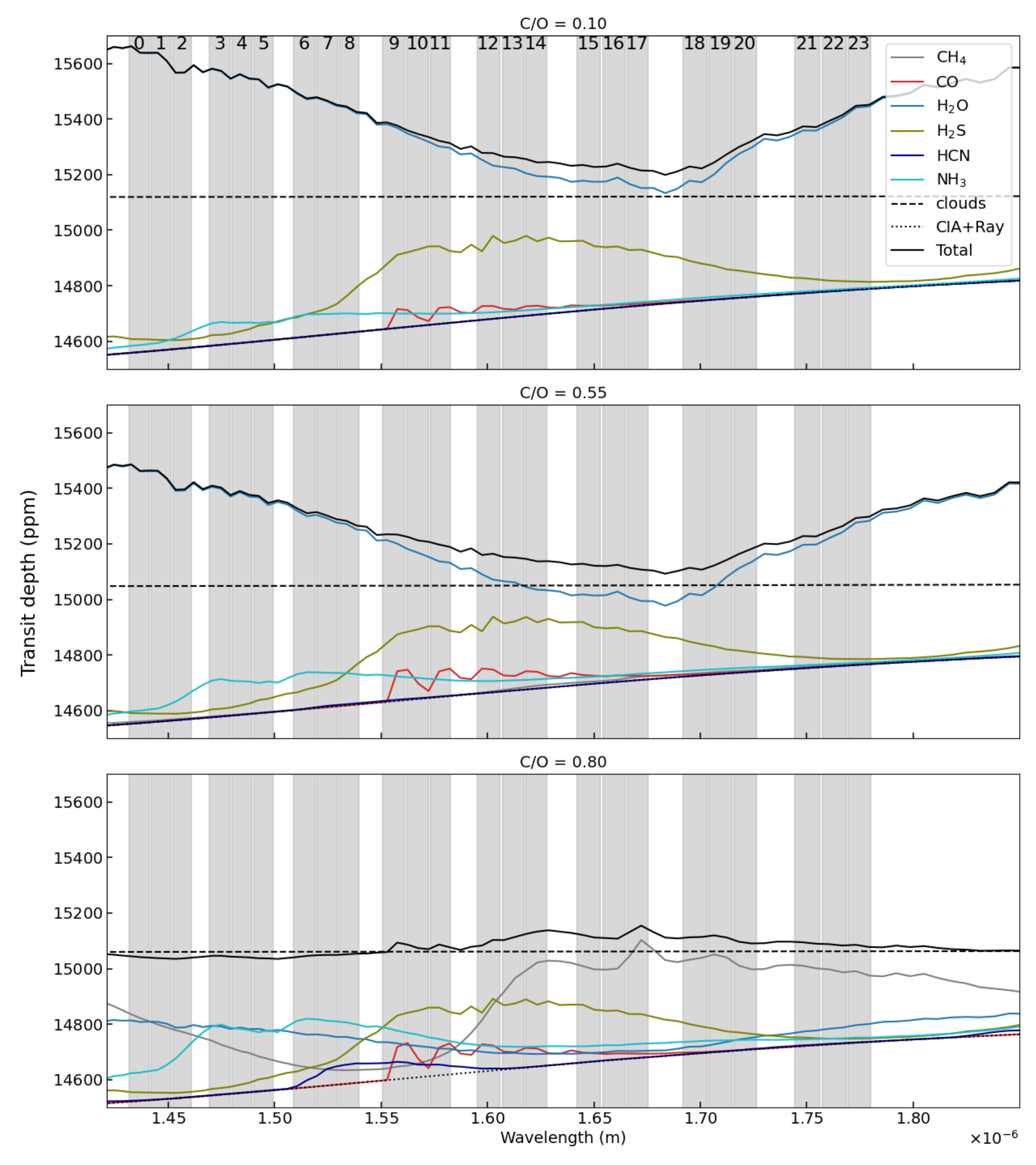}
            \caption{HD~209458~b low-resolution ($\mathcal{R} \approx 500$) transit depth and species contributions simulated with Exo-REM. The grey areas represent the wavelength range of the CRIRES+ orders. On top are indicated the orders index number. From top to bottom: models with a C/O ratio of 0.1, 0.55, and 0.8, respectively. Dotted: the combined contributions of the H$_2$--H$_2$ and H$_2$--He collision-induced absorptions and the effect of Rayleigh scattering. Dashed: the total contribution of the SiO$_2$, Mg$_2$SiO$_3$ and Fe clouds for a cloud coverage of $1.0$. The contributions of CO$_2$, FeH, K, Na, PH$_3$, TiO and VO were negligible in this spectral region and thus are not represented.}
            \label{fig:species_contribution}
        \end{figure*}

        The temperature profile and VMRs obtained are displayed in \autoref{fig:exorem_temperature_profiles} and \autoref{fig:exorem_volume_mixing_ratios}, respectively. The simulated spectra and absorber contributions for each model are displayed in \autoref{fig:species_contribution}.
        
    \subsection{Transmission spectrum model}
        \label{subsec:transmission_spectrum_model}

        \subsubsection{Templates}
            \label{subsubsec:templates}

            \begin{table}
                \centering
                \caption[]{\label{tab:opacity_references} References of the line lists used for our high-resolution spectra}
                \label{tab:line_list_references}
                \begin{tabular}{l l} 
                \hline \hline
                Species     & Reference                             \\
                \hline
                C$_2$H$_2$  & \citet{Rothman2013}                   \\
                CH$_4$      & \citet{Hargreaves2020}                \\
                CO          & \citet{Rothman2010}                   \\
                CO$_2$      & \citet{Rothman2010}                   \\
                H$_2$O      & \citet{Polyansky2018}                   \\
                H$_2$S      & \citet{Rothman2013}                   \\
                HCN         & \citet{Harris2006,Barber2014}         \\
                NH$_3$      & \citet{Coles2019}                     \\
                \hline
                \end{tabular}
            \end{table}
            We model the spectrum of HD~209458~b using the petitRADTRANS\footnote{\url{https://gitlab.com/mauricemolli/petitRADTRANS}} (pRT) package \citep{Molliere2019}. We follow the transmission spectrum construction method of \citet{Blain2024} to obtain our pressure grid, temperature profile (isothermal, with $T = T_\mathrm{eq}$), mass fractions (constant with pressure), and mean molar masses (their step 1). We also use their petitRADTRANS configuration (their step 3), and convolve the spectrum following their step 7 at the expected resolving power of CRIRES+ ($\mathcal{R} = 92\,000$). We also tested templates using $\mathcal{R} = 100\,000$, in order to take into account for the potential 'super-resolution' effect discussed in \autoref{sec:observations}, but found no significant changes in our results. Their other steps, related to Doppler-shifting the lines, are skipped. The references for the line lists used are displayed in \autoref{tab:line_list_references}.

            At first order, CCF analyses are sensitive to the line positions and shapes, but not to their absolute depth. Because of this, and because the template is not deformed in the same way as the observations by the preparing pipeline, this kind of analysis is reliable for species detection, but not for abundance estimation. Hence, the mass fractions of the species in the template model can be set to any value, as long as the lines of the tested species are prominent enough in the modelled spectrum. 
            
            We therefore use the same model basis in all of our templates: the mass fractions of all tested species are set to $-3 \log_{10}(\mathrm{MMR})$, and H$_2$ and He are added so that the sum of MMR is 1, with a He/H$_2$ MMR ratio of 12/37, as in \citet{Blain2024}. We also include the collision-induced absorptions of H$_2$ and He, as well as their Rayleigh scattering effect. We do not include any cloud effect. Then, from this basis, we add the line-by-line ($\mathcal{R} \approx 10^6$) opacities of only the tested species, which we down-sample by a factor of four to reduce memory usage. By cross-correlating these template spectra with the observations, we can thus test for the presence of individual species in the latter.

        \subsubsection{Simulated data}
            \label{subsubsec:simulated_data}
            
            \begin{figure}
               \centering
               \includegraphics[width=\hsize]{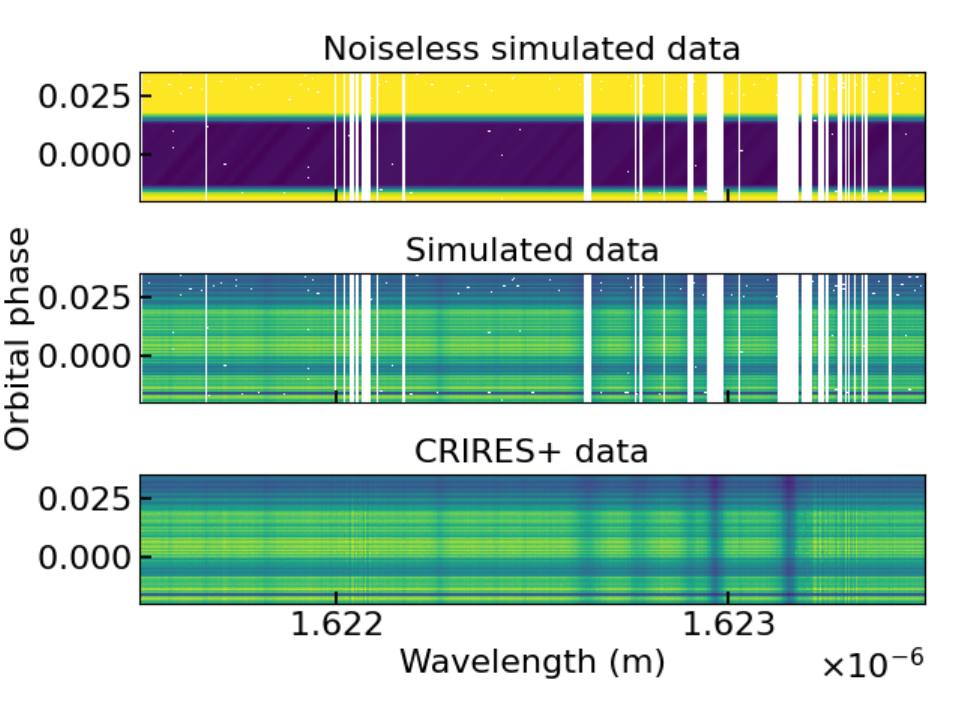}
                  \caption{
                    Simulated data ($Z = 3$, C/O = 0.55, no cloud) and CRIRES+ data of night 3, for a section of order 14. Top row: simulated data with deformation matrix and noise matrix removed. The 'well' around orbital phase 0 is due to the modelled transitting effect. The spectral lines of the planet's atmosphere can be seen within the 'well' as faint dark traces. Middle row: simulated data, including the approximated deformation matrix and the modelled noise matrix. Bottom row: real CRIRES+ data, for comparison. The white pixels are pixels masked during the Polyfit preparing pipeline and the trimming (see \autoref{subsec:preparing_pipeline} of the observations). The spectra are represented in arbitrary units.
                    }
                \label{fig:simulated_data}
            \end{figure}

            We construct a total of nine different sets of simulated data, in a similar manner as the templates. We generate one set for each of our three Exo-REM models (see \autoref{subsec:self_consistent_model}), in which we simulate the clouds by adding opaque layers down to the lowest pressure at which the sum of the cloud opacities reach $1$ in the Exo-REM models (i.e. at $3.9$, $4.4$, and $4.1$ $\log_{10}$(Pa), for a C/O ratio of $0.1$, $0.55$, and $0.8$, respectively). We note that clouds form significantly deeper in our self-consistent Exo-REM models than inferred by \citet{Xue2024} with their free retrieval setup. Because of this, we generate three additional simulated data sets, where we fix the opaque layer lower pressure at $1.7$ $\log_{10}$(Pa) (i.e. 50 Pa). For comparison, we also generate three models with no cloud coverage (opaque layer lower pressure fixed at $7$ $\log_{10}$(Pa)). For the temperature profile, mass fractions\footnote{For the conversion, $X_i = \frac{M_i}{M}V_i$, where $X_i$, $M_i$, and $V_i$ are the mass fraction, molar mass and volume mixing ratio of species $i$, respectively, and $M$ is the mean molar mass of the atmosphere.}, and mean molar masses, we use the values obtained from our Exo-REM models, interpolated to the petitRADTRANS pressure grid. The species C$_2$H$_2$ is not implemented in Exo-REM, so we arbitrarily fix its abundance in all models to $-8$ $\log_{10}$(MMR).
            
            We then modify the spectrum mostly following the simulated data construction procedure described in \citet{Blain2024}, with $V_\mathrm{rest} = -5$ km$\cdot$s$^{-1}$. However, we skip their step 6 bis (the modelling of the telluric transmittances) and modify their step 9 (adding the instrumental deformations) as follows. We run the Polyfit (see \autoref{subsec:preparing_pipeline}) preparing pipeline on the real data, and retrieve the preparing matrix $\mathbf{R}_\mathbf{F}$.  This matrix is the element-wise multiplication of two second-order polynomial fits of the data. From \citet{Blain2024}, if the time- and wavelength-dependent data $\mathbf{F}(t, \lambda)$ can be expressed as $\mathbf{F} = \mathbf{M}_\Theta \circ \mathbf{D} + \mathbf{N}$, where the symbol '$\circ$' represents here the element-wise product \citep{Million2007}, $\mathbf{M}_\Theta$ is an exact model of the planet's spectrum with true parameters $\Theta$, $\mathbf{D}$ is the deformation matrix, and $\mathbf{N}$ is the noise matrix, then the preparing matrix $\mathbf{R}_\mathbf{F}$ is proportional to $\mathbf{1} \oslash \mathbf{D}$. Here the symbol '$\oslash$' represents the element-wise division \citep{Wetzstein2012}. We thus construct the simulated data $\mathbf{F}_{\mathrm{sim}}$ as:
            \begin{eqnarray}
            	\label{eq:simulated_data}
            		\mathbf{F}_{\mathrm{sim}}(t, \lambda) &=& \mathbf{M}_\Theta(t, \lambda) \oslash \mathbf{R}_\mathbf{F}(t, \lambda) + \mathbf{N}(t, \lambda),
            \end{eqnarray}
            where here $\mathbf{M}_\Theta(t, \lambda)$ and $\mathbf{N}(t, \lambda)$ are obtained as in step 10 of \citet{Blain2024}, using the same random seed (12345). The result is displayed in \autoref{fig:simulated_data}.

\section{Methods}
    \label{sec:methods}

    \subsection{Data selection}
        \label{subsec:data_selection}

        \paragraph{Order selection:} The orders 0 to 2 (1.43--1.46 $\mu$m, see \autoref{fig:species_contribution}) of channel A are absent in channel B. For consistency, we remove these three orders from channel A. In addition, orders 3 to 5 (1.46--1.50 $\mu$m) are heavily contaminated by telluric lines, and we decided to discard these orders as well. As a result, we keep on both channel A and B orders 6 to 23, for a total of 18 orders.

        \paragraph{Spectral pixel selection:} Due to the bias subtraction as well as the loss of throughput at the edges of the orders, some spectral pixels on both sides of the orders present negative or near-zero values, that could affect our analysis. In order to avoid this, we discard the first ten and the last five spectral pixels of all orders, exposures, and nights. We thus keep 2033 spectral pixels on each order.

        \paragraph{Bad pixels removal:} After preparation (see \autoref{subsec:preparing_pipeline}), we note the presence of isolated 'cold' and 'hot' pixels, that might be caused by cosmic rays. We also note the presence of 'dashed lines', caused by spectral pixels presenting significantly more variations than their neighbours, at most but not all exposures, and inconsistent with the expected position of telluric or stellar lines (see \autoref{fig:simulated_data}, just after 1.622 $\mu$m). The origin of the latters might be instrumental, or created during the data reduction step. In order to avoid any bias caused by these pixels, we remove them by following the steps below:
        \begin{enumerate}
            \item We fit the observations with a second-order polynomial across wavelengths, for each exposure and each order, then divide the observations with the fit.
            \item We mask the resulting spectra where their medians across exposures are greater than 1.2 and lower than 0.8, in each order and each wavelength.
            \item We mask the 3-$\sigma$ outliers -- determined from the mean and the standard deviation across all orders, exposures, and wavelengths -- of the resulting masked spectra.
            \item We fit the resulting masked spectra with a second-order polynomial across exposures, for each wavelength and each order, then divide them with the fit.
            \item We mask the 3-$\sigma$ outliers -- determined from the mean and the standard deviation across all orders, exposures, and wavelengths -- of the resulting masked spectra.
            \item In each order, we mask a spectral pixel in all the exposures if this spectral pixel was masked in at least 1 percent of the exposures with the previous treatment.
        \end{enumerate}
        The resulting mask is then applied to the unaltered observations (i.e. the data before applying the steps above), before the preparing step (\autoref{subsec:preparing_pipeline}). The spectra resulting from the steps above are otherwise not used. Using this procedure, we increased the $\mathrm{S}\!/\!\mathrm{N}$ metric of our H$_2$O detection from 8.3 to 8.7 $\sigma$ (see \autoref{subesec:H2O_and_kinematics}). On average, considering only our 18 selected orders, and including the preparation step which masks the deepest telluric and stellar lines (see \autoref{subsec:preparing_pipeline}), we mask $\approx 20\%$ of the pixels.


    \subsection{Preparing pipelines}
        \label{subsec:preparing_pipeline}
        As seen in \autoref{fig:simulated_data}, the observations ($\mathbf{F}$) are dominated by telluric and stellar lines, as well as variation of observed flux level, pseudo-continuum, and blaze function ('deformation matrix'). A crucial step of ground-based high-resolution data analysis is to remove as much as possible the deformation matrix in order to be left with only the planet's signal. This is done with algorithms called 'preparing pipelines'.

        \paragraph{Polyfit:} We use Polyfit \citep{Blain2024} to generate mock observations, as described in \autoref{subsubsec:simulated_data}. We follow \citet{Blain2024} and use second-order polynomials to fit both the instrumental deformations, and the telluric and stellar lines. We also mask the prepared data where the fit to the telluric and stellar lines is $< 0.8$. The result of this preparing pipeline are the prepared data 
        \begin{equation}
            \label{eq:polyfit}
            \begin{aligned}
                P_\mathbf{R}(\mathbf{F}) \equiv \mathbf{F} \circ \mathbf{R}_\mathbf{F},
            \end{aligned}
        \end{equation}
        where $\mathbf{R}_\mathbf{F}$ is the preparing matrix corresponding to $\mathbf{F}$. The corresponding uncertainties are $\mathbf{U}_\mathbf{R} \equiv \mathbf{U} \circ \left|\mathbf{R}_\mathbf{F}\right| \circ \sqrt{n}$, where $\mathbf{U}$ are the observations uncertainties, and $n$ is the total variance correction factor of the two fits performed in Polyfit \citep[see][]{Blain2024}.

        \paragraph{SysRem:} We use the SysRem \citep{Tamuz2005} implementation described in Appendix B of \citet{Blain2024}. We allow each passes to run 100 iterations until the convergence criterion ($10^{-15}$) across all orders is reached. Thus, all orders are processed through the same number of iterations, and the same number of passes. The result of this preparing pipeline are the prepared data 
        \begin{equation}
            \label{eq:sysrem}
            \begin{aligned}
                P_{S,n}(\mathbf{F}) \equiv {}^{(n)}\mathbf{F}_S = {}^{(n - 1)}\mathbf{F}_S - \mathbf{a}_n \cdot \mathbf{c}_n,
            \end{aligned}
        \end{equation}
        where $n$ is the number of SysRem passes, ${}^{(0)}\mathbf{F}_S \equiv \mathbf{F} \oslash \overline{\mathbf{X}} - \langle \mathbf{F} \oslash \overline{\mathbf{X}} \rangle_\lambda$ are the observations element-wise divided by their second-order polynomial fits along wavelength ($\overline{\mathbf{X}}$) and mean-subtracted, $\langle \mathbf{F} \oslash \overline{\mathbf{X}} \rangle_\lambda$ is the mean of $\mathbf{F} \oslash \overline{\mathbf{X}}$ along wavelength, and $\mathbf{a}_n$ and $\mathbf{c}_n$ are the vector along exposures and wavelengths, respectively, obtained by SysRem at pass $n$. The corresponding uncertainties are $\mathbf{U}_S \equiv \mathbf{U} \oslash \overline{\mathbf{X}} \circ \sqrt{n_\lambda}$, where $n_\lambda$ is the variance correction factor of the second-order polynomial fit along wavelength \citep[see][]{Blain2024}.

    \subsection{Cross-correlation setup}
        \label{subsec:cross_correlation_setup}

        In order to detect species in the atmosphere of HD~208458~b, we use the cross-correlation technique \citep[e.g.][]{Snellen2010, Brogi2018, Cabot2018, Hawker2018, Alonso2019, Sanchez2019, Sanchez2022}, implemented in the \texttt{redexo} code\footnote{\url{https://github.com/ricolandman/redexo}}. After our data selection step described in \autoref{subsec:data_selection}, we prepare the data using SysRem as described in \autoref{subsec:preparing_pipeline} to obtain the prepared data $P_S(\mathbf{F})(t, \lambda)$. We generate a template $\mathbf{M}$ as in \autoref{subsubsec:templates}. This template is then interpolated on Doppler-shifted wavelengths\footnote{Using the non-relativistic approximation: $\lambda_\mathrm{shift} = \lambda (1 - v / c)$, where $c$ is the speed of light in vacuum.}, from a grid of rest velocities $v$ between $-200$ and $+200$ km$\cdot$s$^{-1}$ with a step of $1$ km$\cdot$s$^{-1}$ (for a total of 401 elements), to obtain the shifted templates $\mathbf{M}_\mathrm{shift}(v, \lambda)$. 
        
        We then cross-correlate this grid of templates with the prepared data following
        \begin{equation}
            \label{eq:ccf}
            \begin{aligned}
                \mathrm{CCF}(t, v) = \sum_i \frac{P_{S,n}(\mathbf{F})(t, \lambda_i) \cdot \left( \mathbf{M}_\mathrm{shift}(v, \lambda_i) - \langle \mathbf{M}_\mathrm{shift} \rangle_\lambda(v) \right)}{\mathbf{U}_S^2(t, \lambda_i)},
            \end{aligned}
        \end{equation}
        where $i$ denotes the wavelength index, and $\langle \mathbf{M}_\mathrm{shift} \rangle_\lambda$ is the mean of $\mathbf{M}_\mathrm{shift}$ along wavelength. The number of SysRem passes used $n$ is discussed in \autoref{subsec:number_of_sysrem_passes}. This process is repeated each night and channel\footnote{There are four nights, three nights have two channels, one night have one channel. Hence, there is a total of seven night/channel entities.} for every order. Within a night/channel, we sum the CCFs of each order to obtain $\mathrm{CCF}_\mathrm{o}(t, v)$. An example of a $\mathrm{CCF}_\mathrm{o}$ map is displayed in \autoref{anx:ccf_o_map}.

        At this point we have one $\mathrm{CCF}_\mathrm{o}(t, v)$ per night and per channel. A good match between the template and the planet's signal in the prepared data leaves a high correlation trace in $\mathrm{CCF}_\mathrm{o}(t, v)$, across exposures. However, this can be difficult to analyse, especially if the correlation with the signal is weak on individual exposure. We can use our knowledge on the expected planet's radial velocity as seen by the observer ($V_\mathrm{obs}(t)$) to ease the analysis. The velocity $V_\mathrm{obs}(t)$ can be written as \citep[e.g.][]{Blain2024}
        \begin{equation}
            \label{eq:kp}
            \begin{aligned}
                V_\mathrm{obs}(t) = v_0 \sin(i_p) \sin\left(2 \pi \phi(t)\right) + V_\mathrm{sys} + V_\mathrm{bary},
            \end{aligned}
        \end{equation}
        where $v_0$ is the planet's orbital velocity assuming a circular orbit, $i_p$ is the planet's orbital inclination, $\phi$ is the orbital phase, $V_\mathrm{sys}$ is the systemic velocity, and $V_\mathrm{bary}$ is the barycentric velocity. We construct a grid of $v_0$ from $-50$ to $+250$ km$\cdot$s$^{-1}$ with a step of $1.3$ km$\cdot$s$^{-1}$ (for a total of 231 elements). This grid is translated into a grid of radial velocity semi-amplitudes ($K_p = v_0 \sin(i_p)$) later, in post-processing. We then shift the rest frame of each $\mathrm{CCF}_\mathrm{o}$ at each exposure from $v$ to $v - V_\mathrm{obs}(t)$, and sum the resulting map along exposures, ignoring out-of-transit exposures, to obtain a shifted cross-correlation map $\mathrm{CCF}_\mathrm{shift}(t, v)$ for each night and each channel.

        To extract detection significance from $\mathrm{CCF}_\mathrm{shift}$, two techniques are commonly used in the literature: the $\mathrm{S}\!/\!\mathrm{N}$ metric \citep[e.g.][]{Brogi2018}, and Welch's $t$-test \citep{Welch1947}. However, \citet{Cabot2018} reported that Welch's $t$-test has a tendency to overestimate detection significance, due to oversampling\footnote{Welch's $t$-test makes the assumption that the two samples are random drawn from normal distributions. In the context of a CCF analysis, this is approximately the case for the out-of-trail sample, but not for the in-trail sample, due to the high correlation of its values. See also \citet{Cheverall2023}.}. Hence, we decided to use the more conservative $\mathrm{S}\!/\!\mathrm{N}$ metric, while it also does not represent an accurate estimation of a species detection confidence \citep[e.g.][]{Cabot2018}. 
        
        We first calculate the sum of $\mathrm{CCF}_\mathrm{shift}$ along exposures and nights/channels to obtain the co-added CCF map ($\mathrm{CCF}_\mathrm{co}$). We then calculate the mean and the standard deviation of elements with a rest velocity outside the $[-15, +15]$ km$\cdot$s$^{-1}$ interval\footnote{Excluding 15 elements centred on 0 km$\cdot$s$^{-1}$, resulting in a sample of $370\ (=401 - 31)$ elements. The choice of this interval is arbitrary.}, that is, the 'out-of-trail' elements of $\mathrm{CCF}_\mathrm{co}$ ($\mathrm{CCF}_\mathrm{oot}$). The signal-to-noise ratio of the CCF is obtained via
        \begin{equation}
            \label{eq:snr}
            \begin{aligned}
                \mathrm{S}\!/\!\mathrm{N}_\mathrm{CCF}(K_p, v) = \frac{\mathrm{CCF}_\mathrm{co}(K_p, v) - \langle \mathrm{CCF}_\mathrm{oot}(K_p, v) \rangle_v}{\mathrm{std}_v(\mathrm{CCF}_\mathrm{oot}(K_p, v))},
            \end{aligned}
        \end{equation}
        where $\mathrm{std}_v(x)$ represents the standard deviation of $x$ along the rest velocities.

    \subsection{Number of SysRem passes}
        \label{subsec:number_of_sysrem_passes}
        \begin{figure}
           \centering
           \includegraphics[width=\hsize]{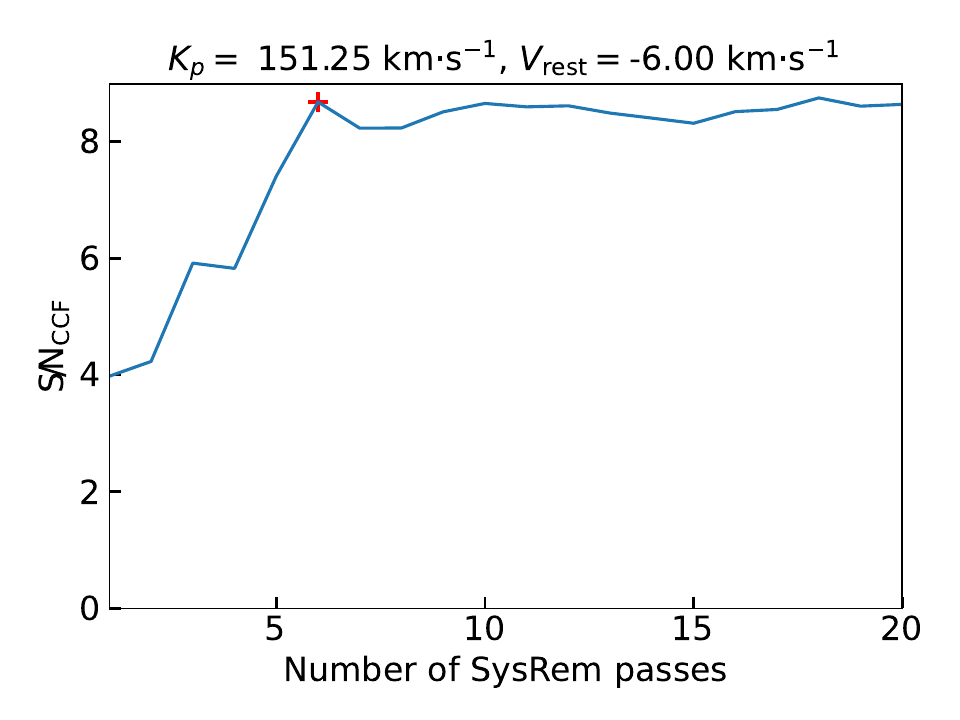}
              \caption{
                    $\mathrm{S}\!/\!\mathrm{N}_\mathrm{CCF}$ ($\sigma$) combining the four nights from the H$_2$O template at $K_p = 151.25$ km$\cdot$s$^{-1}$ and $V_\mathrm{rest} = -6.00$ km$\cdot$s$^{-1}$, with respect to the number of SysRem passes. The red cross indicates the number of SysRem passes with which we chose to perform our analysis.
                }
            \label{fig:sysrem_passes}
        \end{figure}
        
        As described in \autoref{subsec:preparing_pipeline}, the SysRem algorithm works by subtracting systematics from the data. Multiple passes can be applied, to remove 'hidden' linear trends. In principle, instrumental effects, telluric lines and stellar lines should create a limited number of linear trends. Applying too many passes may result in the removal of the planet's signal \citep[e.g.][]{Blain2024}. To estimate the optimal number of SysRem passes to apply, we follow \citet{Alonso2019} and calculate the $\mathrm{S}\!/\!\mathrm{N}$ metric near the expected signal on our $\mathrm{S}\!/\!\mathrm{N}_\mathrm{CCF}$ map for H$_2$O, between 1 and 20 SysRem passes. The result is displayed in \autoref{fig:sysrem_passes}. The $\mathrm{S}\!/\!\mathrm{N}_\mathrm{CCF}$ for H$_2$O increases from approximately 4.0 to 8.7 when the number of SysRem passes is increased from 1 to 6. Further increases in SysRem passes does not significantly improves the $\mathrm{S}\!/\!\mathrm{N}_\mathrm{CCF}$. We decided to be conservative and chose to analyse the data prepared with six SysRem passes.

\section{Results and discussion}
    \label{sec:results_and_discussion}
    
    \begin{figure*}
       \centering
       \includegraphics[width=\hsize]{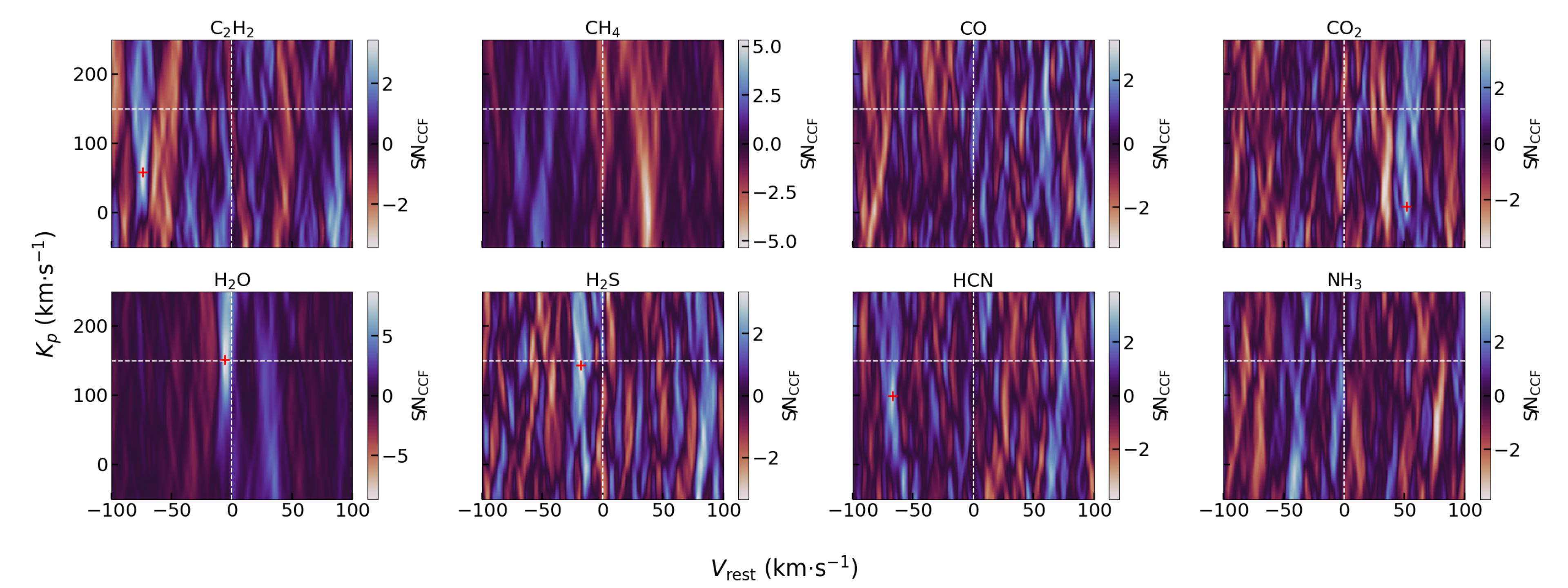}
          \caption{
                CRIRES+ data: cross-correlation signal-to-noise ratio ($\mathrm{S}\!/\!\mathrm{N}_\mathrm{CCF}$) $K_p$--$V_\mathrm{rest}$ maps of all tested species. Negative values indicate an anti-correlation of the template with the data. The vertical and horizontal dashed lines show the expected $K_p$ of the planet and the 0 km$\cdot$s$^{-1}$ rest velocity, respectively. The red cross indicates the location of the maximum $\mathrm{S}\!/\!\mathrm{N}$. Rest velocities lower than $-100$ km$\cdot$s$^{-1}$ and larger than $+100$ km$\cdot$s$^{-1}$ are not shown for clarity. The maximum $\mathrm{S}\!/\!\mathrm{N}_\mathrm{CCF}$ can be in this excluded region, hence the red cross is not on all figures.
            }
        \label{fig:ccf_snr_maps_all_species}
    \end{figure*}

    We display the $\mathrm{S}\!/\!\mathrm{N}_\mathrm{CCF}$ combining the four transits in \autoref{fig:ccf_snr_maps_all_species}. The value and location of the $\mathrm{S}\!/\!\mathrm{N}_\mathrm{CCF}$ peak of each species is displayed in \autoref{tab:ccf_peaks}.
    
    \subsection{H\texorpdfstring{$_2$}~O and kinematics}
        \label{subesec:H2O_and_kinematics}    
        Our H$_2$O $\mathrm{S}\!/\!\mathrm{N}_\mathrm{CCF}$ map is consistent with a planetary signal at $K_p = 151.3^{+31.1}_{-23.4}$  km$\cdot$s$^{-1}$ and $V_\mathrm{rest} = -6^{+1}_{-2}$ km$\cdot$s$^{-1}$. The error bars of these values are obtained by taking the $K_p$ and $V_\mathrm{rest}$ coordinates boundaries of the $\mathrm{S}\!/\!\mathrm{N}_\mathrm{CCF}$ values greater or equal to the maximum $\mathrm{S}\!/\!\mathrm{N}_\mathrm{CCF}$ ($8.7$), minus 1. The radial velocity semi-amplitude we obtain is consistent with the value expected from the literature (see \autoref{tab:general_parameters}): $K_p = 149.4\pm 5.7$ km$\cdot$s$^{-1}$. Results using an alternative line list for water (HITEMP's \citealt{Rothman2010} instead of ExoMol's \citealt{Polyansky2018}) are available in \autoref{anx:H2O_HITEMP}.
        
        The blue-shifted rest velocity we obtain is consistent with the value reported by \citet{Gandhi2019} ($-3.0^{+1.8}_{-1.6}$ km$\cdot$s$^{-1}$) and \citet{Sanchez2019} ($-5.2^{+1.6}_{-1.3}$ km$\cdot$s$^{-1}$), but not consistent with \citet{Snellen2010} ($\approx -2$ km$\cdot$s$^{-1}$), and with \citet{Giacobbe2021}, who did not report any significant Doppler-shift of the planet's spectral features. \citet{Sanchez2019} discuss in details results from General Circulation Models \citep{Rauscher2012, Rauscher2013, Showman2013, Amundsen2016}, concluding that their $V_\mathrm{rest}$ value is in good agreement with these models, except for the one presented in \citet{Rauscher2013}, which includes Ohmic dissipation that reduces wind speeds in the atmosphere. We note that a similar discrepancy in $V_\mathrm{rest}$ in the literature was highlighted for HD~189733~b by \citet{Blain2024}, who suggested that inaccurate data timestamps could be a source of error when estimating this parameter.

    \subsection{Other species}
        We do not observe significant $\mathrm{S}\!/\!\mathrm{N}$ metric peak for any of the other tested species. For C$_2$H$_2$, CH$_4$, CO, CO$_2$, HCN and NH$_3$, the detected $\mathrm{S}\!/\!\mathrm{N}_\mathrm{CCF}$ peak is $\lessapprox 4 \sigma$ and situated at least more than 30 km$\cdot$s$^{-1}$ from the expected planet signal in the $K_p$--$V_\mathrm{rest}$ space (see \autoref{tab:ccf_peaks}).
        
        One species, H$_2$S, show a 'weak' ($\lessapprox 4 \sigma$) peak, but near the expected signal's location. From \autoref{subsec:self_consistent_model}, we expect H$_2$S to be present in HD~209458~b's atmosphere and to have a significant impact on its spectrum. 
        While there is an informal convention in the literature to ignore 'weak' peaks, at face value, a spurious signal of $3.35 \sigma$ or more still has a probability of only $\approx 4\times10^{-4}$ to happen by chance.
        
        To investigate if the peak we detect for H$_2$S could arise from random fluctuations, we assume that our $\mathrm{S}\!/\!\mathrm{N}_\mathrm{CCF}$ maps approximately follow a standard normal distribution, which can be expected from \autoref{eq:snr} if the observation noise is Gaussian. We then calculate the survivor function of the binomial distribution (the regularised incomplete beta function) for the number of $\mathrm{S}\!/\!\mathrm{N}_\mathrm{CCF}$ values $\geq 3 \sigma$, $I_p(k + 1, n - k)$, where $p \approx 1.35 \times 10^{-3}$ is the probability to have $\mathrm{S}\!/\!\mathrm{N}_\mathrm{CCF}$ values $\geq 3 \sigma$ under our assumption, $k$ is the number of $\mathrm{S}\!/\!\mathrm{N}_\mathrm{CCF}$ values $\geq 3$, and $n = 92\,631$ is the number of values in our $\mathrm{S}\!/\!\mathrm{N}_\mathrm{CCF}$ maps. We finally translate the resulting probabilities into $\sigma$-significance. For H$_2$S we have $k = 115$. For comparison, we obtain $k = 2308$ for H$_2$O. The expected $k$ in our case is $\approx 125.04$, translating into a $\sigma$-significance of $-0.85 \sigma$ for H$_2$S. In other words, the number of $\mathrm{S}\!/\!\mathrm{N}_\mathrm{CCF}$ values $\geq 3 \sigma$ for this species is slightly below what is expected for normally distributed samples of the size of our $\mathrm{S}\!/\!\mathrm{N}_\mathrm{CCF}$ map. We conclude that the peak we observe for H$_2$S is, similarly to the other species except H$_2$O, not significant.

        We note that the $\mathrm{S}\!/\!\mathrm{N}_\mathrm{CCF}$ map of CH$_4$ shows a significant anti-correlation peak of $-5.3 \sigma$ at $K_p = 6$ km$\cdot$s$^{-1}$ and $V_\mathrm{rest} = 37$ km$\cdot$s$^{-1}$. This may be the result of remaining residuals. We performed an analysis using our CH$_4$ template on the data prepared with 20 SysRem passes, but only managed to reduce the peak to $-4.9 \sigma$, while the maximum remained both misplaced and $< 3.3 \sigma$. We conclude that, if this peak is caused by residuals, our SysRem implementation was not able to efficiently remove it. While it could be argued that this feature might hamper the detection of CH$_4$ in our data, we note that, according to our results on simulated data, a simultaneous CH$_4$ and H$_2$O detection should be accompanied by at least a CO detection (see \autoref{tab:ccf_peaks_models}). We thus estimate that, even without this feature, it is unlikely that we would have been able to detect CH$_4$.

    \subsection{Comparison with simulated data}
        \label{subsec:comparison_simulated_data}

        \subsubsection{Results}
            \label{subsubsec:simulated_data_results}

            \begin{table}
                \centering
                \caption[]{\label{tab:ccf_peaks} $\mathrm{S}\!/\!\mathrm{N}_\mathrm{CCF}$ peak value and location for the tested species.}
                \begin{tabular}{l l l l l} 
                \hline \hline
                Species     & Value ($\sigma$)	                            & $K_p$ ( km$\cdot$s$^{-1}$)    & $V_\mathrm{rest}$ ( km$\cdot$s$^{-1}$)	\\
                \hline
        		Expected	& --											& $\approx$ 149.6				& $\approx$ 0.0								\\
                C$_2$H$_2$  & 3.44											& 57.8          				& -74.0										\\
                CH$_4$      & 2.81											& 179.8       					& -127.0									\\
                CO          & 3.27											& 39.6          				& 156.0								    	\\
                CO$_2$      & 3.30											& 8.5          				& 52.0								    	\\
                \textbf{H$_2$O}      & \textbf{8.68}											& \textbf{151.3}				          	& \textbf{-6.0}								    	\\
                H$_2$S      & 3.35											& 143.5				          	& -18.0								    	\\
                HCN         & 3.42 											& 99.3							& -67.0								    	\\
                NH$_3$      & 3.86 											& 203.2							& 145.0								    	\\
                \hline
                \end{tabular}
                \tablefoot{
                    The bold font are used to highlight the detected species. Species in normal font are not detected.
                }
            \end{table}

            \begin{table*}
                \centering
                \caption[]{\label{tab:ccf_peaks_models} Simulated data: $\mathrm{S}\!/\!\mathrm{N}$ metric peak value for the tested species in different atmospheric conditions.}
                    \begin{tabular}{l l l l l l l l} 
                    \hline \hline
                    Species	& Data	& C/O = 0.1, cloudy	& C/O = 0.55, cloudy	& C/O = 0.8, cloudy		& C/O = 0.1	& C/O = 0.55	& C/O = 0.8 	\\
                    \hline						
                    CH$_4$ 	& --   	& --				& --  					& --  					& --		& --  			& 12.20			\\
                    CO     	& --   	& --				& --	  				& --  					& --		& 8.25  		& 13.52  		\\
                    CO$_2$ 	& --   	& --				& --					& --					& --		& --			& --			\\
                    H$_2$O 	& 8.68	& 11.86				& 7.25					& --					& 33.84  	& 32.16			& 4.42			\\
                    H$_2$S 	& --  	& --				& --					& --					& 6.10		& 8.96			& 13.75			\\
                    HCN    	& --  	& --				& --					& --					& --		& --			& 4.87			\\
                    NH$_3$ 	& --  	& --				& --					& --					& --		& --			& 4.00			\\
                    \hline
                \end{tabular}
                \tablefoot{
                The 'data' column represents the results obtained from the real CRIRES+ data. The other columns represents results obtained from simulated data. Here 'cloudy' refer to models with $P_c = 1.7$ $\log_{10}$(Pa). Models without this label refer to the cases where $P_c = 7$ $\log_{10}$(Pa). Non-detections, i.e. misplaced peaks or peaks $< 4 \sigma$ are indicated with a '--'. Values for C$_2$H$_2$ are not displayed as this species is never detected due to its artificially low abundance in our models. The value displayed here represent the results of a unique noise realisation and may not be representative of the average expected result.
                }
            \end{table*}
            
            While a CCF analysis such as the one used in this work is inaccurate to estimate chemical abundances and other thermochemical properties (see \autoref{subsubsec:templates}), we can compare our results on the observations to those obtained with simulated datasets. This way, it is possible to check under which conditions detection or non-detection of a given species may be expected with our setup. We display the results for four of our nine simulated datasets in \autoref{fig:ccf_snr_maps_models_all_species}, and for six of them in \autoref{tab:ccf_peaks_models}. We discuss the limits of this approach in \autoref{subsubsec:models_results_summary}. Hereafter we use the term 'no detection' when the maximum of a $\mathrm{S}\!/\!\mathrm{N}_\mathrm{CCF}$ map is lower than $4 \sigma$.
    
            The $\mathrm{S}\!/\!\mathrm{N}_\mathrm{CCF}$ maps we obtain for our simulated data with C/O = 0.1 and $P_c = 1.7$ $\log_{10}$(Pa) are consistent with what we obtain with the observations: a significant H$_2$O signal ($11.9 \sigma$), and no detection otherwise. Results are similar when raising the C/O ratio to 0.55, but with a lower $\mathrm{S}\!/\!\mathrm{N}_\mathrm{CCF}$ peak value ($7.2 \sigma$) for H$_2$O. There is no detection for any species with C/O = 0.8.
    
            When using cloudless models ($P_c = 7$ $\log_{10}$(Pa)), we obtain, as expected, a stronger H$_2$O $\mathrm{S}\!/\!\mathrm{N}_\mathrm{CCF}$ peak with C/O = 0.1 and C/O = 0.55 ($33.8 \sigma$ and $32.2 \sigma$, respectively). By increasing the opaque cloud top pressure, we indeed lessen the cloud effect and increases the lines amplitude in the spectrum. However, H$_2$O is still not detected with C/O = 0.8 (we obtain a peak of $4.4 \sigma$ at $K_p = 40.9$ km$\cdot$s$^{-1}$, but at the expected $K_p$ the peak reaches only $2.8 \sigma$), mainly because of its low abundance, responsible for the low amplitude of its lines in the spectrum. Under these favourable cloud conditions, CH$_4$ and HCN are detected only with C/O = 0.8 ($12.2 \sigma$ and $4.9 \sigma$, respectively), while NH$_3$ is at the limit of detection (peak at $4.0 \sigma$). CO is detected only for C/O $\geq$ 0.55. H$_2$S is detected at all the tested C/O ratios. On the other hand, C$_2$H$_2$ and CO$_2$ are not detected in any simulated dataset. This result is expected for C$_2$H$_2$, given its artificially low abundance in our models, and for CO$_2$, given its absence of significant spectral absorption in the CRIRES+ wavelength range observed in this study\footnote{We note that assuming a bulk C/O ratio $> 1$, C$_2$H$_2$ and HCN are expected to be the major C-bearing species behind CO, and above CH$_4$, see e.g. \citet{Molliere2015}. These species may thus be detectable in band H at these higher C/O ratios.}.
    
            Our results with the Exo-REM cloud models ($P_c \approx 4$ $\log_{10}$(Pa)) show strong similarities with our results with the cloudless simulated data. The modelled opaque cloud layer is indeed located at too high pressures to significantly affect the spectrum. Similarly to the cloudless models, they do not reproduce the observations well. In addition, as mentioned in \autoref{subsubsec:simulated_data}, this would also be inconsistent with the cloud pressure inferred by \citet{Xue2024}. This suggests that the Exo-REM cloud model we used cannot accurately simulate cloud formation on HD~209458~b, or that different condensing species than the one implemented, or hazes, are involved.
    
            These models also show that our non-detection of CO or CO$_2$ in band H is not inconsistent with the detection of these species in other works \citep{Snellen2010, Brogi2019, Gandhi2019, Xue2024}. For CO, the reported detections were from band K (around $2.3$ $\mu$m) observations, and the lowest reported abundance is from \citet{Brogi2017}, with $-3.80^{+0.51}_{-0.53}$ $\log_{10}$(VMR), which is fully consistent with the  $\log_{10}$(VMR) of our Exo-REM model with C/O = 0.1 and $Z$ = 3 ($\approx -3.6$). For CO$_2$, the detection \citep{Xue2024} is inferred from its $v3$ band around 4.4 $\mu$m, and is consistent with a C/O ratio of 0.1.

            It is worth noting that higher C/O ratios seem to favour H$_2$S detection. While at constant metallicity, the abundance of H$_2$S does not vary significantly with the C/O ratio (see \autoref{fig:exorem_volume_mixing_ratios}, variations are caused by the slight changes in the temperature profile, affecting the chemical balance), the decrease in H$_2$O abundance makes the H$_2$S lines more prominent in the spectrum (see \autoref{fig:species_contribution}).

            \begin{table*}
            	\centering
            	\caption[]{\label{tab:studies_comparison} Comparison of this work with a selection of other HD~209458~b studies.}
            		\begin{tabular}{l l l l l} 
            		\hline \hline
												& \citet{Hawker2018}			& \citet{Giacobbe2021}								& \citet{Xue2024} 				& This work				\\
            		\hline															
            		Instrument					& VLT/CRIRES   	 				& TNG/GIANO-B 										& \textit{JWST}/NIRCam 					& VLT/CRIRES+   		\\
            		$\mathcal{R}$				& $\approx$100$\,$000			& $\approx$50$\,$000								& $\approx$1500 				& $\approx$92$\,$000 	\\
            		$\lambda$ range ($\mu$m)	& 2.29--2.35, 3.18--3.27		& 0.95--2.45										& 2.36--4.02, 3.86--5.06		& 1.51--1.78   		    \\
            		Observations				& 4 2$^{\mathrm{ry}}$ transits            				& 5 transits										& 2 transits\tablefootmark{a} 	& 4 transits  		    \\
            		Method						& CCF analysis					& CCF analysis										& Free retrieval 				& CCF analysis			\\
					Detected species			& CO, H$_2$O, HCN				& C$_2$H$_2$, CH$_4$, CO, H$_2$O, HCN, NH$_3$		& CO$_2$, H$_2$O 				& H$_2$O				\\
                    C/O							& $\gtrapprox 1$\tablefootmark{b}& $\gtrapprox 1$					                & $0.11^{+0.12}_{-0.06}$			& $\approx$ 0.1--0.55\tablefootmark{b}	\\
                    $\log_{10}(Z)$				& --				          & $\approx -1$ to 1 									& $0.54^{+0.30}_{-0.23}$		& $\approx$ 0.5 to 1\tablefootmark{b}		\\
					$P_c$ ($\log_{10}($Pa$)$)	& --							& $\approx-0.5$\tablefootmark{b, c}	& $1.69^{+0.50}_{-0.69}$		& $\approx$ 1.7\tablefootmark{b}~		\\  
					\hline
            		\end{tabular}
            		\tablefoot{
                        \tablefoottext{a}{One transit for each wavelength range.}
            			\tablefoottext{b}{Limited or no parameter space exploration.}
            			\tablefoottext{c}{Cloud fraction of 0.4.}
            		}
            \end{table*}

        \subsubsection{Possibility for a simultaneous detection of H\texorpdfstring{$_2$}~O and HCN in band H} 
            Interestingly, even under the very favourable conditions of a cloudless atmosphere and a 'perfect' template, we are unable to detect both HCN and H$_2$O with $Z = 3$ and a bulk C/O = 0.8 -- effectively corresponding to C/O $\approx$ 1 above $10^{3}$ Pa (see \autoref{subsec:self_consistent_model}). HCN (and even more so NH$_3$) are less than $1 \sigma$ from the detection limit. Due to Exo-REM limitations, we did not investigate if the HCN (and NH$_3$) $\mathrm{S}\!/\!\mathrm{N}_\mathrm{CCF}$ would increase with a higher C/O ratio at this metallicity, but, in all cases, such an increase would make a H$_2$O detection even less likely, due to its removal by Si-bearing condensates (as discussed in \autoref{subsec:self_consistent_model}). We note that \citet{Giacobbe2021} reported that their chemical model does not take this effect into account. On the other hand, a lower Si/O ratio would allow for more H$_2$O to remain above the cloud level, but to our knowledge there is no formation model considering this scenario in the literature. We also tested if increasing the metallicity to $Z = 10$ could allow for this simultaneous detection, but this was not the case (see \autoref{anx:co0.8_z10}). In this scenario, only a $12\%$ change in H$_2$O abundance would have resulted in the complete removal (hence, undetectability) of this species. By staying in the upper atmosphere however, it seems to act as a screening for the HCN features, which as a consequence is no more detected.
            
            Thus, it seems, according to our current knowledge and to our setup, that a simultaneous detection of HCN and H$_2$O would indeed require a C/O ratio fine-tuned to an arguably extreme level, on top of requiring a cloudless (or an almost clear) atmosphere, which seems inconsistent with the observations \citep[e.g,][]{Sing2016, Pinhas2018, Giacobbe2021, Xue2024}. However, an argument could be made that the terminator may have different C/O ratios: for example, one can imagine a hotter evening terminator where Si-bearing clouds did not condense, lowering in this region the effective (gas phase) C/O ratio, and increasing the H$_2$O signal strength, while being at a high enough C/O in the gas phase at the colder morning terminator, due to nightside cloud condensation, such that carbon species such as CH$_4$, HCN and C$_2$H$_2$ are at an increased abundance \citep[see][for more details on this scenario]{Sanchez2022}. We did not properly test this scenario in this work, but the 3-D model including cloud formation of HD~209458~b from \citep{Lines2018} do not predict an asymmetric cloud coverage along longitudes. According to a 2-D simulation, a relative chemical uniformity is expected across the planet \citep{Tsai2024}. A chemically asymmetric terminator scenario for HD~209458~b seems thus difficult to explain with current state-of-the-art models.
    
            As previously mentioned, two previous works claimed a simultaneous detection of H$_2$O and HCN in HD~209458b: \citet{Hawker2018} using VLT/CRIRES data and \citet{Giacobbe2021} using Telescopio Nazionale Galileo(TNG)/GIANO-B data, both at a high resolution. We note that \citet{Giacobbe2021} performed their CCF analysis only on the orders for which, for a given species, the detection of an injected signal was estimated sufficient. Several works \citep[e.g.][]{Debras2023, Blain2024}, especially \citet{Cheverall2023}, shown that this kind of procedure is likely to introduce biases in the analysis. Indeed, $\mathrm{S}\!/\!\mathrm{N}_\mathrm{CCF}$ maps contains random fluctuations due to the cross-correlation of the template, mostly with noise. It is thus likely that this translates sometimes into positive correlations near the expected position in $K_p$--$V_\mathrm{rest}$ space. By selecting orders that favours the detection strength of an injected signal, there is a risk for a spurious signal to emerge from the combination of these favourable fluctuations. \citet{Hawker2018}, instead, optimised the number of their Principal Component Analysis (PCA) preparing pipeline passes on each order for each species, as to maximise the detection of an injected signal. It is possible that this method also introduces biases by leaving in the prepared data systematics that would create spurious signals, although this specific case was not investigated in \citet{Cheverall2023}. On the other hand, the wavelength range of \citet{Hawker2018} data (3.18--3.27 $\mu$m) is more favourable to HCN detection than ours (1.51--1.78 $\mu$m) and those of \citet{Giacobbe2021} (0.95--2.45 $\mu$m), due to the proximity of the strong HCN $v1$ fundamental band around 3.02 $\mu$m \citep[e.g.][]{Harris2006, Barber2014}. In all cases, a re-analysis of these data without a data optimisation method would be useful to understand this discrepancy.
        
        \subsubsection{Summary}
            \label{subsubsec:models_results_summary}
            
            Our model with $Z = 3$, C/O = 0.1 and $P_c = 1.7$ $\log_{10}$(Pa) reproduces well our observations, although with a higher maximum $\mathrm{S}\!/\!\mathrm{N}_\mathrm{CCF}$. It appears clearly though that the strength of the H$_2$O signal in the $\mathrm{S}\!/\!\mathrm{N}_\mathrm{CCF}$ map can be decreased both by decreasing the pressure of the opaque cloud top layer, or by increasing the C/O ratio. We thus estimate that, according to our models, a higher C/O ratio and a higher opaque cloud top layer pressure are consistent with our observations, to a certain degree. With a C/O ratio of $\approx 0.8$, we are not able to reproduce the observations, under any cloud condition. C/O ratios $< 0.1$ and $P_c < 1.7$ $\log_{10}$(Pa) could also be possible. This is consistent with the results obtained by \citet{Xue2024}, but not consistent with those inferred by \citet{Hawker2018} and \citet{Giacobbe2021}. A comparison of our results with these studies is displayed in \autoref{tab:studies_comparison}. 
            
            We note that a lower maximum $\mathrm{S}\!/\!\mathrm{N}_\mathrm{CCF}$ than expected could also arise from imperfect line lists (see \autoref{anx:H2O_HITEMP}), or residuals not captured by our simulated datasets. We note also that this conclusion comes from a limited parameter space exploration. Finally, these results are only valid for the unique noise realisation we tested. The $\mathrm{S}\!/\!\mathrm{N}$ metric on those simulated data are expected to vary around their mean value with a standard deviation of 1, hence all species, especially those at the limit of detection, could be detected or undetected depending on whether or not the noise realisation is favourable.
                   
\section{Conclusion}
    \label{sec:conclusion}
    
    We analysed CRIRES+ data of four HD~209458~b transits, covering wavelengths from $1.51$ to $1.78$ $\mu$m at a resolving power of $\approx 92\,000$, and prepared with SysRem \citep{Tamuz2005}. Combining the four nights, our CCF analysis detects a H$_2$O signal compatible with HD~209458~b at $8.68 \sigma$. We tested the data with C$_2$H$_2$, CH$_4$, CO, CO$_2$, H$_2$S, HCN, and NH$_3$, but did not detect any of these species.

    In complement, we compared the above results with nine simulated datasets, using the temperature profiles and abundances from our self-consistent atmospheric model Exo-REM, spectra generated with petitRADTRANS, and deformations and uncertainties extracted from the observations. We found that our $\mathrm{S}\!/\!\mathrm{N}_\mathrm{CCF}$ maps are consistent with a simulated dataset corresponding to $Z = 3$, C/O = 0.1 and $P_c = 1.7$ $\log_{10}$(Pa). However, the observations are also consistent, at this metallicity, with a combination of higher C/O ratios and higher opaque cloud top pressures, but not with cloudless atmospheres with a C/O ratio $\geq 0.55$, and not under any cloud condition with a C/O ratio of 0.8. Indeed, increasing the C/O ratio decreases the strength of the H$_2$O signal, due to chemical process favouring the formation of CO when more carbon is available, and due to the removal of H$_2$O because of the condensation of Si-bearing species. In contrast, increasing the opaque cloud top pressure increases the signal of all species, due to the removal of the cloud's spectral dampening effect.

    With these models, we also show that a simultaneous detection of HCN and H$_2$O in HD~209458~b's atmosphere with CRIRES+ in band H, even when using a very specific C/O value, seems unlikely. This would require a different HCN chemical model -- or a different cloud formation model -- than the one we used, or strong regional chemical discrepancies in HD~209458~b, which does not seem to be expected \citep{Lines2018, Tsai2024}. Our non-detection of any species except H$_2$O is not consistent with the detection of C$_2$H$_2$, CH$_4$, HCN and NH$_3$ by \citet{Giacobbe2021} from GIANO-B data. Instead, our results are consistent with those obtained by \citet{Xue2024}, from \textit{JWST} observations.

\begin{acknowledgements}
    The authors thank A. Sánchez-López for his precious help with the observation setup. This research has made use of the NASA Exoplanet Archive, which is operated by the California Institute of Technology, under contract with the National Aeronautics and Space Administration under the Exoplanet Exploration Program. This research has made use of the Exoplanet Follow-up Observation Program (ExoFOP; DOI: 10.26134/ExoFOP5) website, which is operated by the California Institute of Technology, under contract with the National Aeronautics and Space Administration under the Exoplanet Exploration Program.
\end{acknowledgements}

\bibliographystyle{aa} 
\bibliography{bibliography} 

\begin{appendix}
    \section{Example of a \texorpdfstring{$\mathrm{CCF}_\mathrm{o}$}~ map}
        \label{anx:ccf_o_map}
        
        \begin{figure}
           \centering
           \includegraphics[width=\hsize]{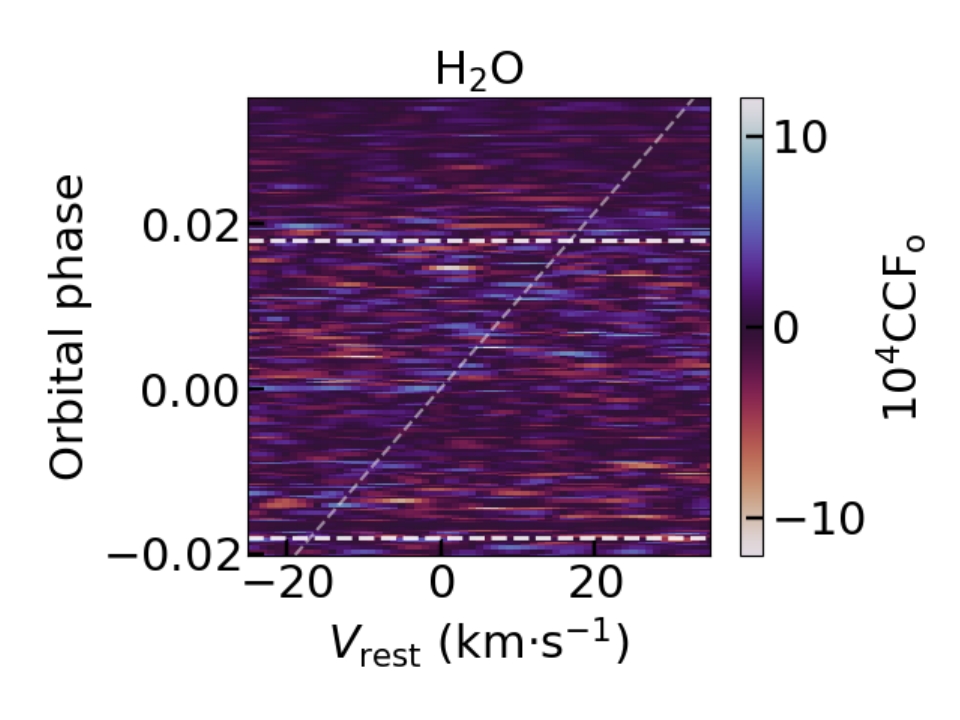}
              \caption{
                    CRIRES+ data: $\mathrm{CCF}_\mathrm{o}$ map for night 3, channel B, using the H$_2$O template. The horizontal white lines represent the orbital phases of the HD~209458~b's expected transit ingress (bottom) and egress (top). The slanted line represent HD~209458~b's expected radial velocities, corrected from the barycentric velocities, with $V_\mathrm{rest} = 0$ km$\cdot$s$^{-1}$. Rest velocities lower than $-25$ km$\cdot$s$^{-1}$ and larger than $+35$ km$\cdot$s$^{-1}$ are not shown for clarity.
                }
            \label{fig:ccf_map_vr_op_n3B_H2O}
        \end{figure}

        \begin{figure}
           \centering
           \includegraphics[width=\hsize]{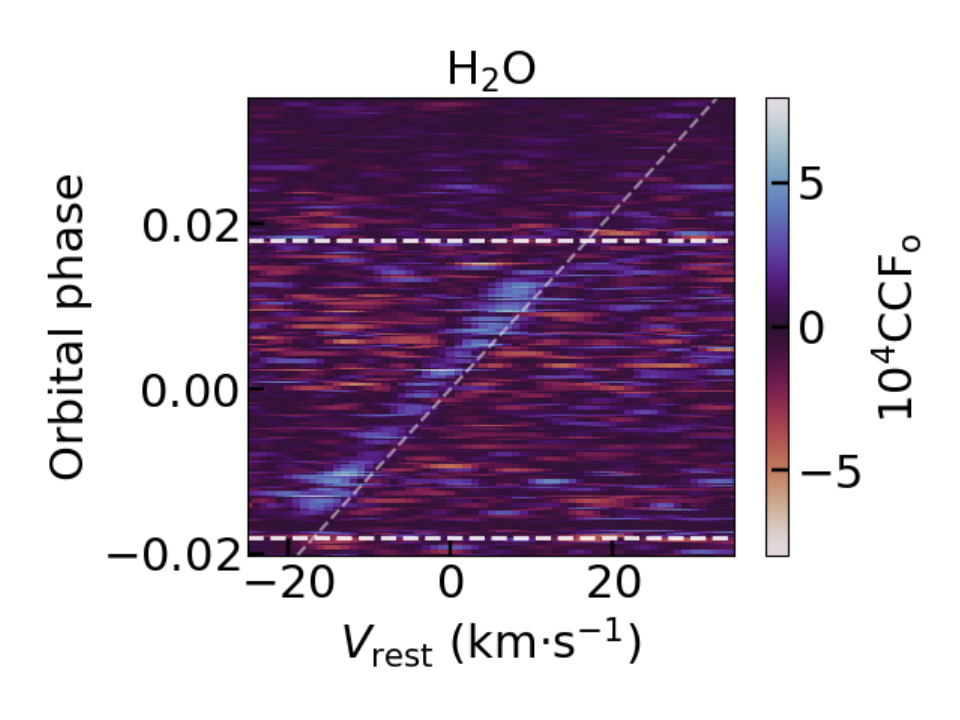}
              \caption{
                    Simulated data: $\mathrm{CCF}_\mathrm{o}$ map for night 3, channel B, using the H$_2$O template. The model has $Z = 3$, C/O = 0.1 and no cloud. The horizontal white lines represent the orbital phases of the HD~209458~b's expected transit ingress (bottom) and egress (top). The slanted line represent HD~209458~b's expected radial velocities, corrected from the barycentric velocities, with $V_\mathrm{rest} = 0$ km$\cdot$s$^{-1}$. Rest velocities lower than $-25$ km$\cdot$s$^{-1}$ and larger than $+35$ km$\cdot$s$^{-1}$ are not shown for clarity.
                }
            \label{fig:ccf_map_vr_op_n3B_H2O_model}
        \end{figure}

        In \autoref{fig:ccf_map_vr_op_n3B_H2O} we represent the $\mathrm{CCF}_o$ map obtained on the observations for night 3, channel B, using the H$_2$O template. For comparison, we display the same map obtained with simulated data in \autoref{fig:ccf_map_vr_op_n3B_H2O_model}, assuming a cloudless atmosphere.
        
    \section{Results using the H\texorpdfstring{$_2$}~O HITEMP line list}
        \label{anx:H2O_HITEMP}
        \begin{figure}
           \centering
           \includegraphics[width=\hsize]{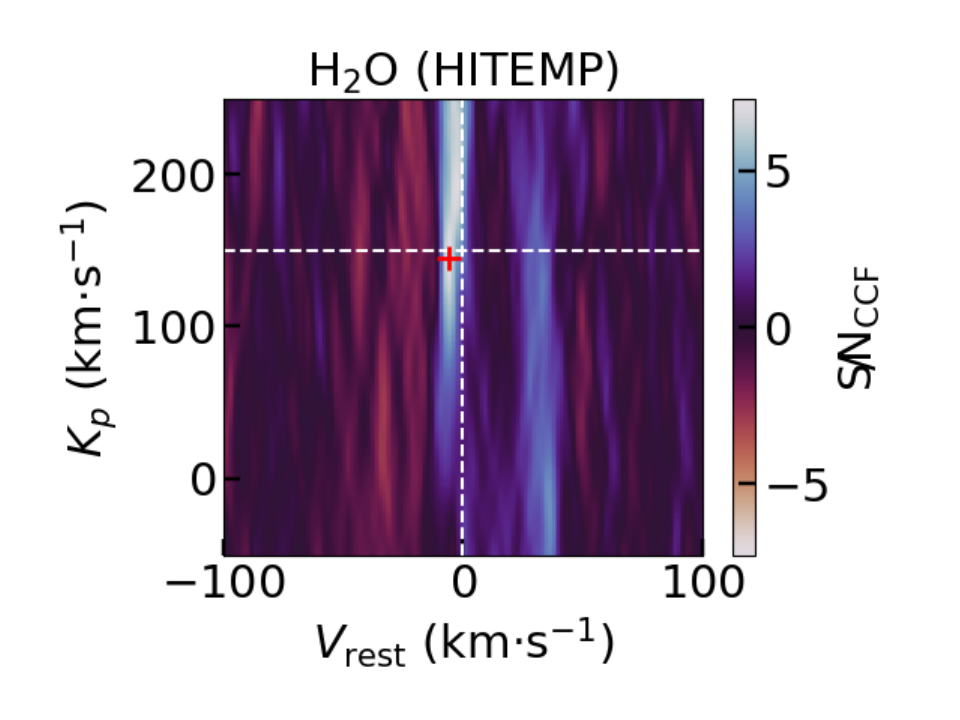}
              \caption{
                CRIRES+ data: cross-correlation signal-to-noise ratio ($\mathrm{S}\!/\!\mathrm{N}_\mathrm{CCF}$) $K_p$--$V_\mathrm{rest}$ maps using a H$_2$O template with the HITEMP line list. Negative values indicate an anti-correlation of the template with the data. The vertical and horizontal dashed lines show the expected $K_p$ of the planet and the 0 km$\cdot$s$^{-1}$ rest velocity, respectively. The red cross indicates the emplacement of the maximum $\mathrm{S}\!/\!\mathrm{N}$. Rest velocities lower than $-100$ km$\cdot$s$^{-1}$ and larger than $+100$ km$\cdot$s$^{-1}$ are not shown for clarity.
                }
            \label{fig:ccf_snr_map_H2O_HITEMP}
        \end{figure}

        \begin{figure*}
           \centering
           \includegraphics[width=\hsize]{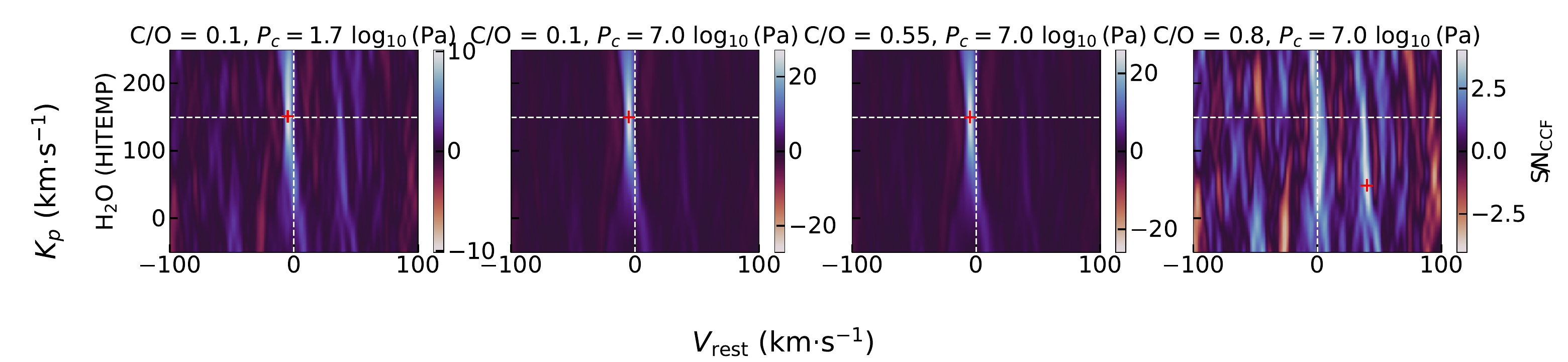}
              \caption{
                    Simulated data: cross-correlation signal-to-noise ratio ($\mathrm{S}\!/\!\mathrm{N}_\mathrm{CCF}$) $K_p$--$V_\mathrm{rest}$ maps using a H$_2$O template with the HITEMP line list for a selection of simulated datasets constructed with the POKAZATEL H$_2$O line list. The leftmost column displays the results obtained with a C/O ratio of 0.1, and an opaque cloud top pressure of 1.7 $\log_{10}$(Pa). Going to the right, the columns display the results from cloudless ($P_c = 7$ $\log_{10}$(Pa)) simulated datasets, with C/O ratio of 0.1, 0.55, and 0.8, respectively. Negative values indicate an anti-correlation of the template with the data. The vertical and horizontal dashed lines show the expected $K_p$ of the planet and the 0 km$\cdot$s$^{-1}$ rest velocity, respectively. The red cross indicates the location of the maximum $\mathrm{S}\!/\!\mathrm{N}$. Rest velocities lower than $-100$ km$\cdot$s$^{-1}$ and larger than $+100$ km$\cdot$s$^{-1}$ are not shown for clarity.
                }
            \label{fig:ccf_snr_maps_models_H2O_HITEMP}
            \end{figure*}

        In addition to our main analysis using the POKAZATEL \citep[][see \autoref{tab:line_list_references}]{Polyansky2018} H$_2$O line list, we also performed an analysis using the H$_2$O line list from the HITEMP database \citep{Rothman2010}. The $\mathrm{S}\!/\!\mathrm{N}_\mathrm{CCF}$ maps for the observations and a selection of simulated data are displayed in \autoref{fig:ccf_snr_map_H2O_HITEMP} and \autoref{fig:ccf_snr_maps_models_H2O_HITEMP}, respectively.

         We observe a $\mathrm{S}\!/\!\mathrm{N}_\mathrm{CCF}$ peak of $7.3 \sigma$ at $K_p = 145.0^{+103}_{-1}$ km$\cdot$s$^{-1}$ and $V_\mathrm{rest} = -6^{+5}_{-1}$ km$\cdot$s$^{-1}$. The constraints and peak value are thus slightly less significant than using the POKAZATEL line list, but the results are otherwise similar.

         We performed the same analysis on our simulated data, using a H$_2$O HITEMP template on a spectra generated with the POKAZATEL line list. This results as expected in a significantly decreased $\mathrm{S}\!/\!\mathrm{N}_\mathrm{CCF}$ peak for the cloudless case ($\approx -6 \sigma$ compared to the POKAZTEL-on-POKAZATEL results). On the cloudy case, the effect is less impactful with a loss of only $\approx -2 \sigma$.

    \section{Results on simulated data with \texorpdfstring{$Z = 3$}~}
        \label{anx:simulated_data_z3}
        \begin{figure*}
           \centering
           \includegraphics[width=\hsize]{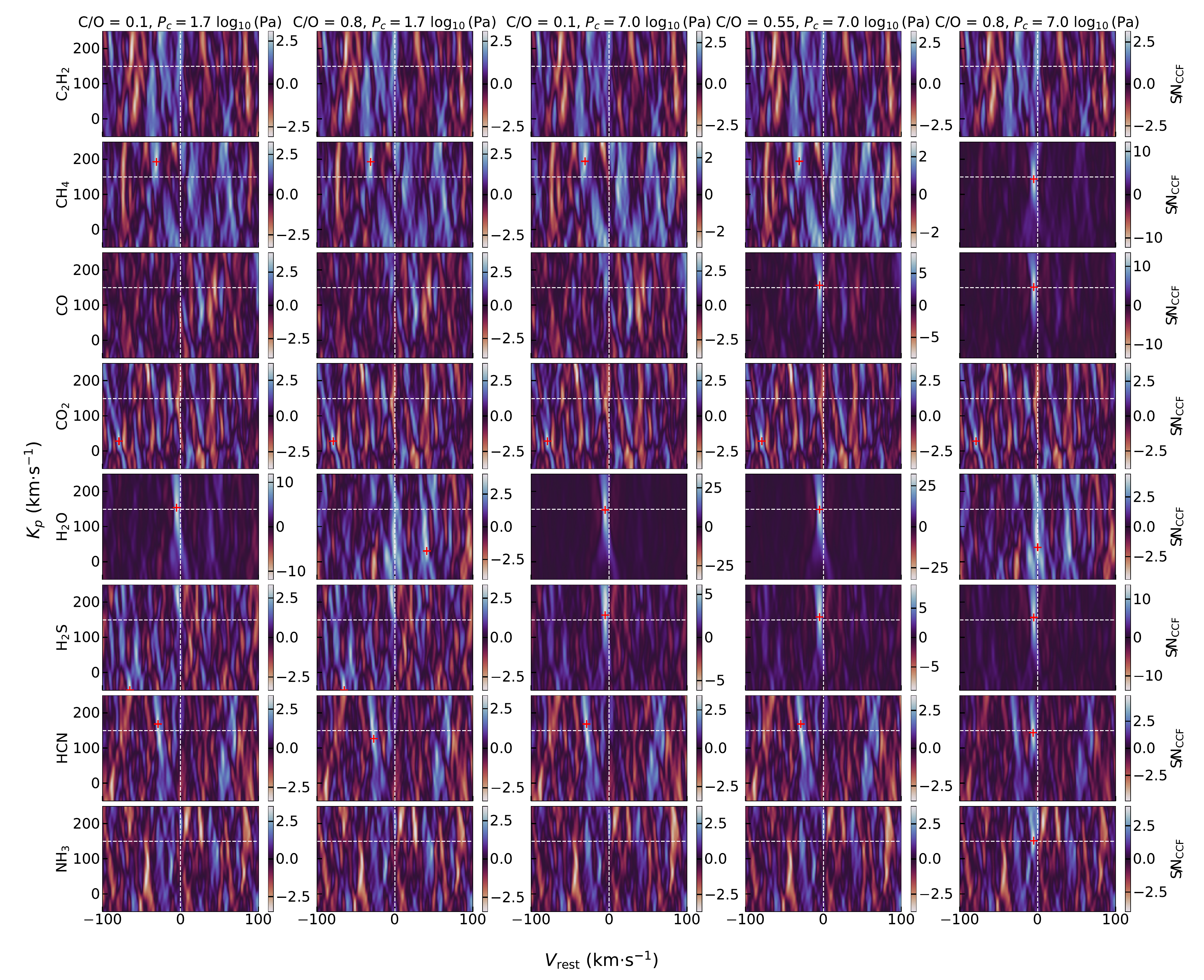}
              \caption{
                    Simulated data: cross-correlation signal-to-noise ratio ($\mathrm{S}\!/\!\mathrm{N}_\mathrm{CCF}$) $K_p$--$V_\mathrm{rest}$ maps of all tested species. Each row displays the map corresponding to a different species. The leftmost column displays the results obtained with a C/O ratio of 0.1, and an opaque cloud top pressure of 1.7 $\log_{10}$(Pa). Going to the right, the columns display the results from cloudless ($P_c = 7$ $\log_{10}$(Pa)) simulated datasets, with C/O ratio of 0.1, 0.55, and 0.8, respectively. Negative values indicate an anti-correlation of the template with the data. The vertical and horizontal dashed lines show the expected $K_p$ of the planet and the 0 km$\cdot$s$^{-1}$ rest velocity, respectively. The red cross indicates the location of the maximum $\mathrm{S}\!/\!\mathrm{N}$. Rest velocities lower than $-100$ km$\cdot$s$^{-1}$ and larger than $+100$ km$\cdot$s$^{-1}$ are not shown for clarity. The maximum $\mathrm{S}\!/\!\mathrm{N}_\mathrm{CCF}$ can be in this excluded region, hence the red cross is not on all figures. All simulated data use the same noise seed, hence random features are the same on different models.
                }
            \label{fig:ccf_snr_maps_models_all_species}
        \end{figure*}

        In \autoref{fig:ccf_snr_maps_models_all_species} we display the $\mathrm{S}\!/\!\mathrm{N}_\mathrm{CCF}$ maps of four of our nine simulated datasets.

    \section{Results on simulated data with C/O = 0.8 and \texorpdfstring{$Z = 10$}~}
        \label{anx:co0.8_z10}
        \begin{figure*}
           \centering
           \includegraphics[width=\hsize]{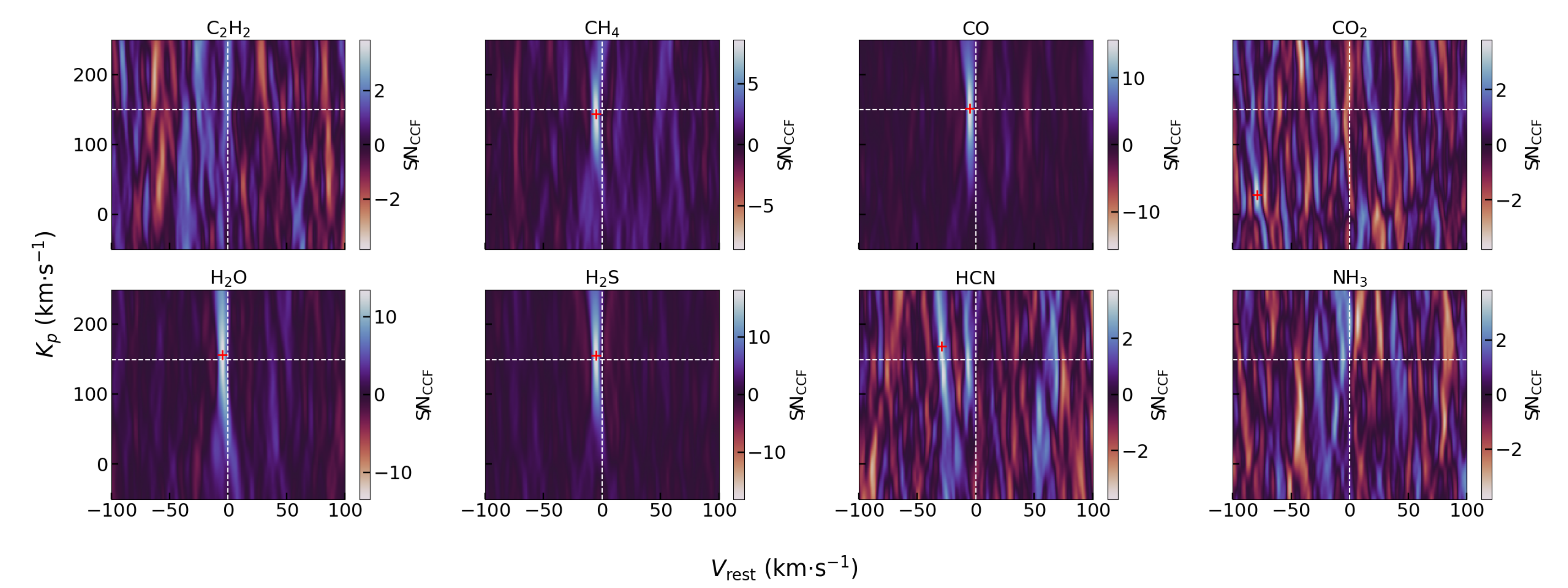}
              \caption{
                    Simulated data: cross-correlation signal-to-noise ratio ($\mathrm{S}\!/\!\mathrm{N}_\mathrm{CCF}$) $K_p$--$V_\mathrm{rest}$ maps of all tested species, using a simulated dataset with C/O = 0.8 and $Z = 10$, and no cloud ( $P_c = 7$ $\log_{10}$(Pa)). Negative values indicate an anti-correlation of the template with the data. The vertical and horizontal dashed lines show the expected $K_p$ of the planet and the 0 km$\cdot$s$^{-1}$ rest velocity, respectively. The red cross indicates the location of the maximum $\mathrm{S}\!/\!\mathrm{N}$. Rest velocities lower than $-100$ km$\cdot$s$^{-1}$ and larger than $+100$ km$\cdot$s$^{-1}$ are not shown for clarity. The maximum $\mathrm{S}\!/\!\mathrm{N}_\mathrm{CCF}$ can be in this excluded region, hence the red cross is not on all figures.
                }
            \label{fig:ccf_snr_maps_model_co0.8_z10_all_species}
        \end{figure*}

        In complement to our simulated data at $Z = 3$, we also tested if a cloudless model at $Z = 10$ and C/O = 0.8 would allow for a simultaneous detection of H$_2$O and HCN. The results are displayed in \autoref{fig:ccf_snr_maps_model_co0.8_z10_all_species}.

        The $\mathrm{S}\!/\!\mathrm{N}_\mathrm{CCF}$ in this scenario are similar to those obtained with at C/O = 0.8 and $Z = 3$. HCN is this time undetected (the peak is misplaced and at $3.7 \sigma$), but H$_2$O is detected with a peak at $13.5 \sigma$. NH$_3$ is also undetected, with a misplaced peak at $3.8 \sigma$.
        
        By increasing the metallicity, we increase the amount of absorbing gases in the atmosphere, hence the temperature, which rises at all layers by $\approx 50$ K. This slightly shifts the H$_2$O--SiO chemical balance so that H$_2$O ends up slightly more abundant ($\approx 12 \%$) than SiO at the SiO$_2$ condensation layer, preventing the complete removal of H$_2$O from the atmosphere. The overall higher temperatures also favour the formation of CO and CO$_2$ at the detriment of CH$_4$. With less of the latter available, and despite a more metallic atmosphere and a more favourable kinematic balance, the HCN abundance only increases by $\approx 15 \%$ compared to the $Z = 3$ case, below the CH$_4$/HCN quench pressure at $\approx 4.8 \log_{10}$(Pa).

\end{appendix}
    
\end{document}